\documentclass[aps,prb,twocolumn,letterpaper]{revtex4}
\usepackage{graphicx}
\usepackage{dcolumn}
\usepackage{amsmath}
\usepackage{amssymb}
\usepackage[usenames,dvipsnames]{color}
\usepackage{float}
\usepackage{epstopdf}

\def\rv{{\bf r}}

\def\Av{{\bf A}}
\def\kv{{\bf k}}

\def\beq{\begin{equation}}
\def\eeq{\end{equation}}
\def\beqa{\begin{eqnarray}}
\def\eeqa{\end{eqnarray}}

\begin{document}

\title{Dirac cones, Floquet side bands and theory of time resolved ARPES}
\author{Aaron Farrell, A.~Arsenault and T.~Pereg-Barnea}
\affiliation{Department of Physics and the Centre for Physics of Materials, McGill University, Montreal, Quebec,
Canada H3A 2T8}
\date{\today}
\begin{abstract}
%
%
Pump-probe techniques with high temporal resolution allow one to drive a system of interest out of equilibrium and at the same time, probe its properties.  Recent advances in these techniques open the door to studying new, non-equilibrium phenomena such as Floquet topological insulators and superconductors.  These advances also necessitate the development of theoretical tools for understanding the experimental findings and predicting new ones.   
In the present work, we provide a theoretical foundation to understand the non-equilibrium behaviour of a Dirac system. We present detailed numerical calculations and simple analytic results for the time evolution of a Dirac system irradiated by light. These results are framed by appealing to the recently revitalized notion of sidebands\cite{Farrelltop,Farrelltriv}, extended to the case of non-periodic drive where the fast oscillations are modified by an envelope function.  We apply this formalism to the case of photocurrent generated by a second, probe pulse.  We find that, under the application of circularly polarized light, a Dirac point only ever splits into two copies of sidebands. Meanwhile, the application of linearly polarized light leaves the Dirac point intact while producing side bands.  In both cases the population of the side bands are time dependent through their non-linear dependence on the envelope of the pump pulse. Our immediate interest in this work is in connection to time and angle resolved photoemission experiments, where we find excellent qualitative agreement between our results and those in the literature.\cite{Wang} However, our results are general and may prove useful beyond this particular application and should be relevant to other pump-probe experiments.
\end{abstract}
\maketitle
\section{Introduction}
One of the greatest triumphs in the last decade of condensed matter research has been the theoretical prediction\cite{Bernevig, moore,fu} and subsequent experimental realization\cite{hsieh, konig, Roth} of the topological insulator (TI). These materials are insulating in the bulk, while their edge plays host to topologically protected metallic modes with energies lying in the band gap of bulk states. The existence of these edge-states makes TIs of great fundamental and practical interest with applications ranging from quantum computation to spintronics.  Moreover, the discovery of topological systems leads to a new classification of possible states of matter.

While many of the topological systems can be understood by non-interacting, clean systems at equilibrium the study of topological states is not limited to those.  The effects of disorder, for example may drive a system in and out of a topological state.\cite{li09,borchmann} It is therefore interesting to ask whether there is a knob that can be tuned to alter the topological properties of a system.
 One auspicious route towards the generation of a TI comes from considering time-periodic perturbations\cite{Lindner, Gu,Oka,Usaj, Calvo,Torres,Wang, Leon, Rudner, Kitagawa2,Kundu2,Katan, Jiang, Kundu, Liu, Wu, Wang2, Delplace, Li4, Dehghani1,Dehghani2, paraj, Farrelltop, Farrelltriv}. In these systems, a time-periodic perturbation, is applied to a topologically trivial system and drives into a non-equilibrium topological state. As continuous time translational invariance is broken, it is no longer appropriate to discuss energy eigenstates. One must instead talk about their quasi-energy spectrum, which is the closest analogue to an energy spectrum for a system with discrete time-translational invariance\cite{Sambe}. The topological state created with an external, time-periodic perturbation is called a Floquet topological insulator (FTI) and it exhibits edge-states in the gap of its quasi-energy spectrum\cite{Lindner}. 

The notion of a FTI has garnered much attention lately, and has enjoyed experimental validation in the field of photonic crystals, where Floquet states can be simulated in the laboratory\cite{rechtsman2013photonic}. However, a solid state verification of a Floquet topological state and several issues regarding feasibility have been raised\cite{sentef2015theory} . The first of these is that most available periodic perturbations are not perfectly periodic, but have an envelope function in addition to the periodic signal. The second, perhaps more pressing, issue has to do with the experimentally available frequencies. Present discussions in the literature are valid in the large frequency limit $\Omega\gg \Omega_{BW}$, $\Omega$ being the applied frequency and $\Omega_{BW}$ being the frequency of the band-width of the system. However, available technology in terahertz is sub-bandwidth. This small frequency is believed to be problematic as it will lead to a complicated quasi-energy structure which may obscure any potential topological effects.

Given the above complications, our goal in the present Paper is to understand the behaviour of a topological system in the presence of a non-periodic and sub-bandwidth external perturbation, while probing the system continuously over time. We will work with Dirac cone dispersion, typical for a three dimensional topological insulator surface, in order to develop a fundamental understanding and will not discuss a Floquet topological insulator at this point. Our work is inspired by measurements of Wang {\it et al} in Ref.~[\onlinecite{Wang}]. This group used time resolved-angle resolve photoemission spectroscopy (TR-ARPES) to view the evolution of surface states of Bi$_2$Se$_3$, a three dimensional topological insulator. We find excellent qualitative agreement with these results.

We employ the language of sidebands recently used in Refs.~[\onlinecite{Farrelltop, Farrelltriv}]. This language allows us to develop the following physical picture of the time-dependent system.  The pump pulse excites the system out of equilibrium.  Its time dependence is generally composed of fast oscillations modified by a slow envelope function.  The fast oscillations normalize the band dispersion and produces side band copies.  The side bands are populated statistically with weights which depend approximately as Bessel functions on the ratio pump pulse amplitude to its frequency.  Since the pump amplitude is time dependent through its envelope the side band weights are also time dependent.  This time dependence allows the system to interpolate between its equilibrium state at the distant past to the Floquet/side band picture when the pump is applied. 
	
The above picture leads to the three main results of this work. The first of these is that even though applying a sub-bandwidth perturbation to a system may ``fold" many states into the Floquet zone, only a few of these states have any spectral weight and contribute to physical processes. In the present example, we consider a Dirac cone, which has effectively an infinite band-width. We find that only states within a couple $\hbar\Omega$ from the Fermi surface have any statistical weight in our side-band picture. 
%
Second, we work in a regime where the time scale over which the pump pulse envelope is changing is much longer than the period time of the drive.  In this regime we develop simple, analytic expressions. In other cases (such as a quench) the same formalism can be applied and solved numerically.  It should be noted that when the pump electric field is turned off, the system does not necessarily relax immediately to its equilibrium state.  This is particularly clear in the case of the physical gauge we adopt, as explained in Appendix~\ref{ap:Gauge} and may lead to interesting effects like persistent Hall response\cite{Wilson}. 
Finally, we show that the structure of a Dirac cone colludes with circularly polarized light to produce {\em only two} sidebands for momenta near the Dirac point. This is quite remarkable; the spectral weight of the equilibrium Dirac point states is entirely shared between two sidebands. It therefore behaves as two massive Dirac points, with different masses. These two cones share the spectral weight of the original Dirac cone, and the weights are found analytically as a function of time. 

The intuition developed here, as well as the satisfactory results in view of recent experiments\cite{Wang}, will add to a current ongoing discussion in the literature regarding the stability of Floquet-states\cite{DAlessio,DAlessio2, karthik, PhysRevE.90.012110, Ponte}. Our side-band interpretation in concert with an understanding of Floquet states and the results of Ref.~[\onlinecite{Wang}] provide an intuitive physical picture of the side bands and their probabilities. 

The rest of this paper is organized as follows. In the following section we discuss some fundamentals of the Floquet formalism in order to introduce the side-band intuition of Refs.~[\onlinecite{Farrelltop,Farrelltriv}]. We move on to present our model and methods. In Section \ref{sec:results} we present our results and discussion for two polarizations of light.  The appendices detail various technical aspects of the work. 

\section{Preliminaries}
\subsection{Sidebands}
We begin with a brief discussion of Floquet theory as it pertains to the language of sidebands. Consider the time-dependent Schr\"odinger equation
\beq
i\hbar\partial_t|\psi(t)\rangle = H(t) |\psi(t)\rangle
\eeq
where $H(t+T)=H(t)$ is a Hamiltonian with period $T$. Defining $\Omega=2\pi/T$ the principle result of Floquet theory is that the steady states of the above system can be written as\cite{Sambe}
\beq
 |\psi(t)\rangle= e^{-i \eta t/\hbar}  |\phi(t)\rangle
\eeq
where $(H(t)-i\hbar \partial_t) |\phi(t)\rangle= \eta  |\phi(t)\rangle$ and $ |\phi(t+T)\rangle= |\phi(t)\rangle $. The eigenvalues $\eta$ are typically called the quasi-energies. The quasi-energies are only unique up to integer multiples of $\hbar\Omega$, as can be seen by noting that $e^{in\Omega t} |\phi(t)\rangle$ is an eigenvalue of $(H(t)-i\hbar \partial_t)$ with quasi-energy $\eta + n\hbar\Omega$ and also meets the boundary condition $|\phi(t+T)\rangle= |\phi(t)\rangle$. Thus all quasi-energies are defined within a first "Floquet zone", an interval of energies of width $\hbar\Omega$. The center of this zone is, of course, arbitrary. The quasi-energy spectrum in the first Floquet zone can be copied at integervals of $\hbar\Omega$ above and below to generate the full quasi-energy spectrum. 

We now introduce the side-bands. Since $ |\phi(t)\rangle$ is periodic, we are free to express it as a discrete Fourier series $ |\phi(t)\rangle=\sum_n e^{-in\Omega t} |n\rangle $.  The full wave function reads
\beq\label{sidebandsWF}
 |\psi(t)\rangle=\sum_n e^{-i(\eta+n\hbar\Omega) t/\hbar} |n\rangle  
\eeq
The states $|n\rangle$ are determined by solving the eigenvalue equation $\sum_m \left(H_{n-m}-n\hbar\Omega\delta_{n,m}\right) |m\rangle = \eta |n\rangle$ where $H_n=\int_0^T \frac{dt}{T} e^{in\Omega t} H(t)$. 

The intuitive picture we wish to take away from Eq.~(\ref{sidebandsWF}) is the following. In a time periodic system the steady states are a linear combination of definite energy states with energies $\eta + n\hbar\Omega$ and probability $\langle n|n\rangle$. This follows from either inspecting Eq.~(\ref{sidebandsWF}) or by noting that the average energy over one cycle of the period reads
\beq
\bar{E} =\int_0^T \frac{dt}{T} \langle \psi(t)|H(t)|\psi(t)\rangle = \sum_n \langle n|n\rangle (\eta+n\hbar\Omega)
\eeq

For pedagogical reasons we no consider the application of the above theory to a time-independent system. To be more concrete, let's say we have an applied perturbation with frequency $\Omega$ but a vanishingly small amplitude. In this limit $H_n=\delta_{n,0}H$ and the eigenvalue equation becomes $(H-n\hbar\Omega)|n\rangle = \eta |n\rangle$. The solution to this system is $|n\rangle = \delta_{N,n} |\zeta\rangle$ with $\eta=E-N\hbar\Omega$, where $H|\zeta\rangle = E|\zeta\rangle$ and $N$ is an integer that takes $E$ and moves it into the first Floquet zone we have chosen for our problem. Thus when the time periodic fields are turned off the system is, of course, found in eigenstates of the static Hamiltonian. These eigenstates can, of course, be defined in a first Floquet zone, but if this first Floquet zone does not contain $E$ (i.e. if $N\ne0$ in the language above) then there exists a quasi-energy $E-N\hbar\Omega$, but this state has {\em zero} probability of being occupied in the first Floquet zone because $\langle n|n\rangle=\delta_{n,N}$. One must move to the $N^{\text{th}}$ Floquet zone where this state is occupied with unit probability. The morale of this exercise is the following. 
When working in the Floquet zone the quasienergy spectrum might be dense with folded bands.  However, the 'occupation' of a given state (its weight in the time dependent wavefunction) maybe zero in the first Floquet zone, leaving only a few relevant states.

Starting from the above limit, as we turn on the time dependence there are two effects that take place. First, electrons beginning in the original eigenstates develop some probability to absorb or emit photons and thus their unit probability of being found in one Floquet zone gets smeared into other, adjacent Floquet zones. This creates ``copies" of the original band structure analogous to those proposed by Tien and Gordon several decades ago\cite{TienGordon}. Second, unlike the physics of Ref.~[\onlinecite{TienGordon}], in our present system these side-bands can also be modified in a non-trivial way. This occurs when states corresponding to absorbing/emitting different number of photons hybridize. This can lead to important effects such as gaps in these side-bands opening.  

\begin{figure*}[]
  \setlength{\unitlength}{1mm}

   \includegraphics[scale=.35]{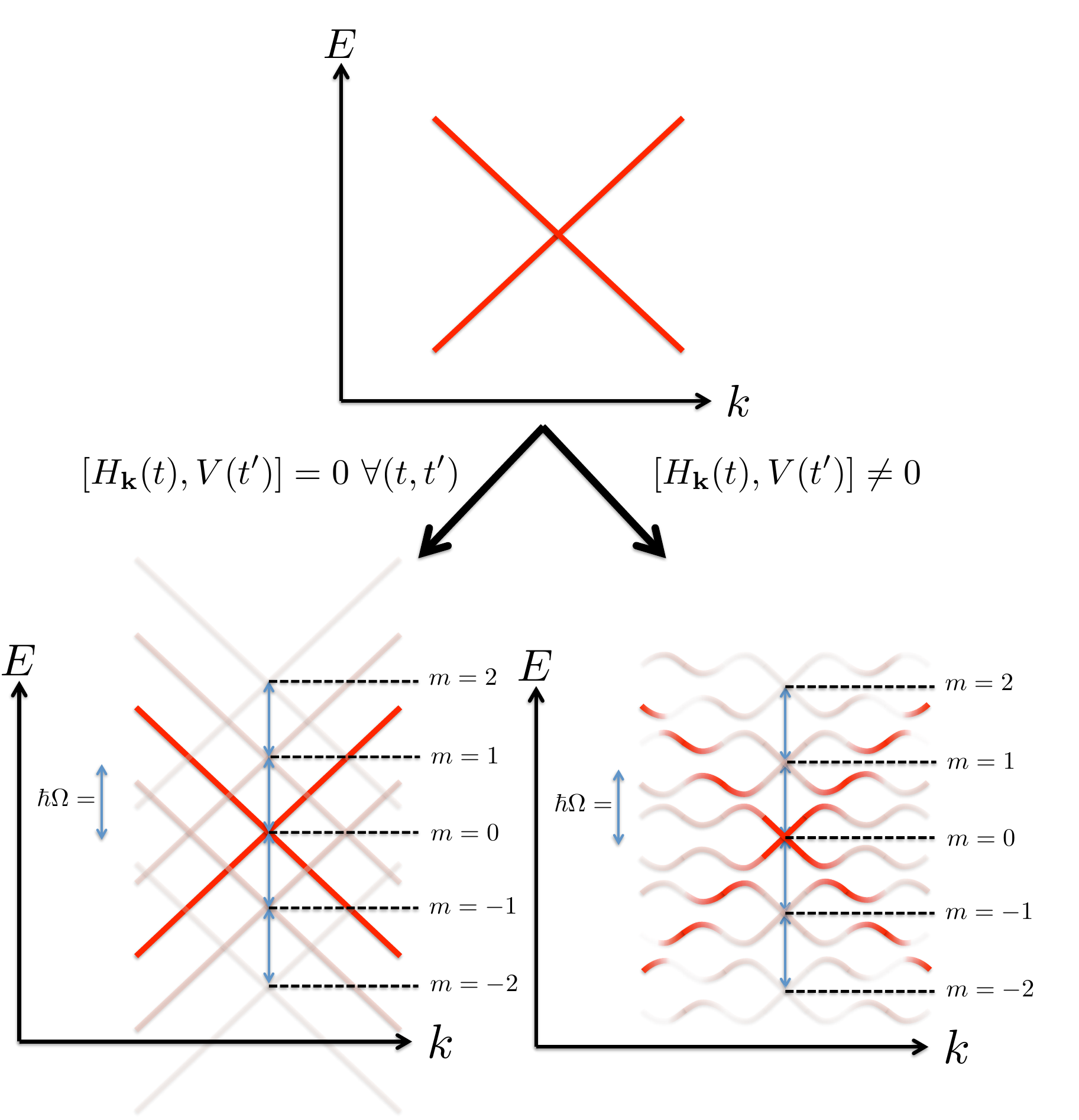}
\caption{{\small
Schematic picture of the main results of this paper. The original Dirac cone is split into side bands, with side bands further away from the original cone receiving less "weight". In the figure this is signified using lighter colours for less probable sidebands. Now, if the operator describing the time periodic field, $V(t)$, commutes with the original Hamiltonian, $H_{\kv}(t)$, then this splitting is all that happens. If these two operators do not commute sidebands hybridize and the band structure becomes modified by, e.g., having gaps opened.  
   }
     }\label{Cartoon}
\end{figure*}

The above interpretation is important when applying a probe of well defines energy to a time periodic system.  An example for this situation is studies in Refs.~[\onlinecite{Farrelltop, Farrelltriv}] where we calculate the transport properties of periodically driven quantum well heterostructures.  Namely, the edge-states in these systems, whether naturally occurring or driven, are split into side-bands. As a result, certain transport signatures of these edge states, for example $2e^2/h$ conductance, are fragmented. In Refs.~[\onlinecite{Farrelltop, Farrelltriv}] we have discussed how a sum rule\cite{Kundu} can be used to salvage these transport signatures. This sum rule is rooted in the understanding that systems in a time-periodic field have their energy bands modified by the time-periodic perturbation and also that these bands are split into side-bands. Crucially, these side-bands are only occupied with a certain probability, and, for reasonable field strengths, this probability decreases with the separation in energy between the original energy eigenvalue and the side-band eigenvalue that we're interested in. Thus it is usually appropriate to treat only eigenstates within several multiples of $\hbar\Omega$ from the Fermi level.

The above observations are important to keep in mind when applying Floquet theory to look at the quasi-energies by themselves. When the energy scale $\hbar\Omega$ is small compared to the band-width of the equilibrium model, the quasi-energy spectrum becomes very convoluted as {\em many} eigenstates are ``folded" back into the Floquet zone. Making predictions based on this spectrum alone then becomes an arduous task. The discussion above, and the results to follow, illustrate that one must keep in mind that even though the quasi-energy spectrum may become complicated in this limit, only quasi-energies resulting from folding of energies within a few $\hbar\Omega$ of the Fermi energy contribute significantly to observables. The information about these probabilities is contained in the often ignored side-band states $|n\rangle$ and their statistical weight.

Our model is a generic Dirac cone and no cut-off is considered, thus our effective band-width is infinite. We subject this system to terahertz frequency light $\hbar\Omega\sim 30$meV.  Looking only at the quasi-energy spectrum of this system the Dirac cone will be folded back into the Floquet zone infinitely many times and would thus be meaningless. We therefore approach the system in a slightly different manner, while keeping in mind the side-band language discussed above. Provided that the field is turned on slowly compared to the frequency of the light, the system evolves into a state described by a splitting of its original bands into side-bands. In cases where the operator describing the external field commutes with the static Hamiltonian {\it at all times}, this side-band splitting is the only effect of the light, i.e. we see no hybridization and no gap opening. In all other cases there are additional modifications of the side bands. In either case, we see that for physical field strengths only the first couple of side-bands carry any spectral weight in these simulations, in spite of the fact that the system is subjected to low-frequency light. These central results of our work are summarized in the schematic in Fig.~\ref{Cartoon}. This intuition should be relevant to related experiments on time-dependent systems and will be crucial in driving a topological state with externally applied light. 

\begin{figure*}[]
  \setlength{\unitlength}{1mm}

   \includegraphics[scale=.45]{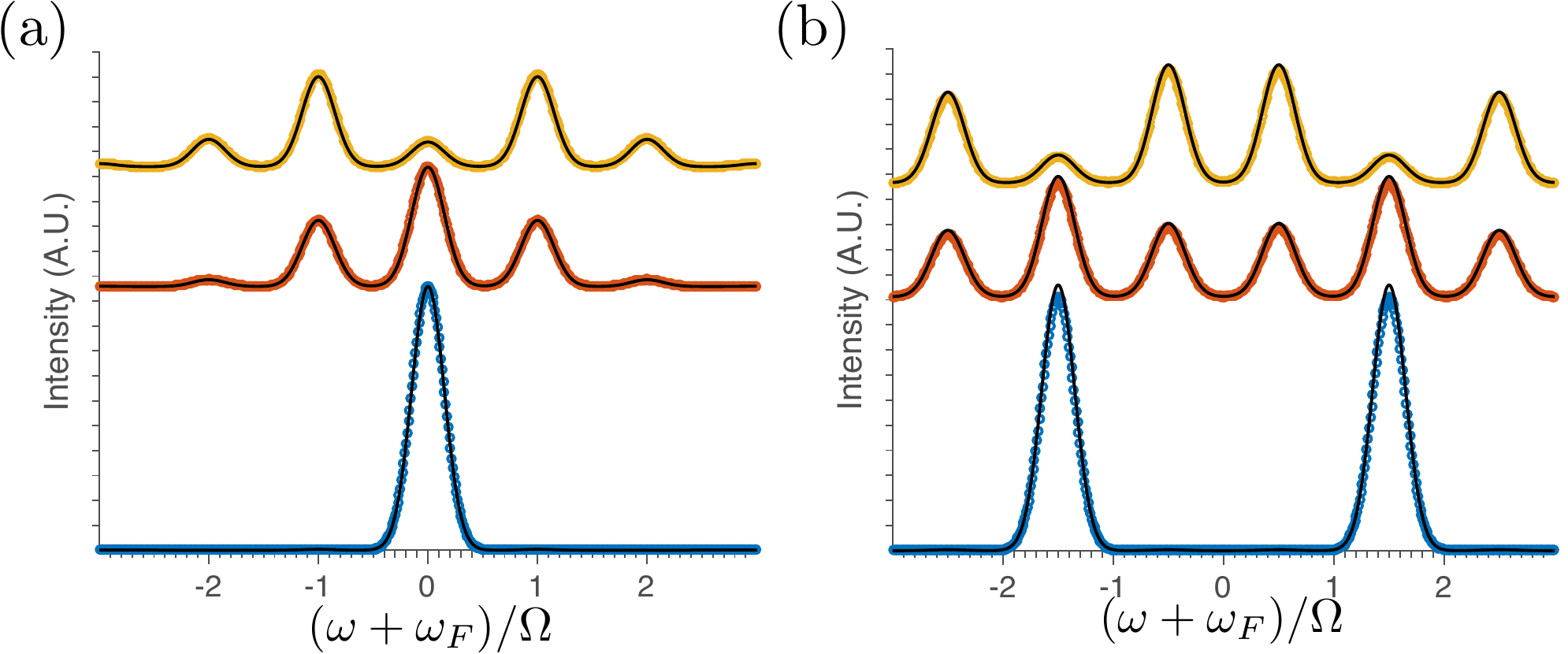}
\caption{{\small Comparison of numerical results found by integrating the time dependent Dirac equation and the analytic approximation in Eq.~(\ref{Linearfinal}) for $I(k_x,0,\omega,t_\mathcal{O})$. The left plots $I(0,0,\omega,t_\mathcal{O})$ for different delay times $t_\mathcal{O}$ while the right plots $I(0.05\AA^{-1},0,\omega,t_\mathcal{O})$ also for different delay times. In these plots the bottom plot is for $t_\mathcal{O}=-500$fs, the middle for $t_\mathcal{O}=-100$fs and the top is for $t_\mathcal{O}=0$fs. In all plots the solid line is the approximation in  Eq.~(\ref{Linearfinal}) while the circles are numerical results. There is excellent agreement between the numerics and our approximation for all three delay times. 
     }
     }\label{fig:linearcompare}
\end{figure*}

\subsection{Model Hamiltonian}
We begin with the following Dirac Hamiltonian
\beq
h_\kv =\hbar v_F( \kv\times \vec{\sigma} )\cdot \hat{z}-\mu \sigma_0
\eeq
where $v_F$ is the Fermi velocity, $\vec{\sigma}_i=\sigma_i$ is a vector of Pauli matrices and $\mu$ the Fermi energy. The above Hamiltonian is immediately applicable to the surface of a three dimensional topological insulator (TI) and should also be relevant to graphene in the limit where any applied field doesn't induce intervalley scattering. 

We now envisage the above system irradiated by an electromagnetic field. To keep our theoretical model simple we assume this field is spatially constant over the sample size. This should be approximately true for the terahertz type radiation considered here where the wavelength of the light should be tens of microns\cite{Wang}. We model this electromagnetic field as follows
\beq
{\bf E}_{pump}(t) = E_0 e^{-\frac{t^2}{2T_{pump}^2}} {\bf E}_{\Omega}(t)
\eeq
where $E_0$ is the amplitude of this pump pulse, $T_{pump}$ is the width of the pulse and ${\bf E}_{\Omega}(t)$ is the monochromatic component of the field. In this paper we consider two scenario's: (1) Linearly polarized light, with ${\bf E}_{\Omega}(t)=\sin{\Omega t} \hat{x}$ and (2) circularly polarized light in which case ${\bf E}_{\Omega}(t)= \sin{\Omega t} \hat{x}-\cos{\Omega t} \hat{y}$.

We introduce the above field via minimal coupling, ignoring the Zeeman effect, as we expect the dominant contribution to come from the electron's orbital motion. We choose a Gauge such that the electric scalar potential $\Phi=0$ and ${\bf E}_{pump}(t) =-\partial_t \Av_{pump}(t)$, see appendix C for more details. Thus we have $\Av_{pump}(t) = - \int_{-\infty}^t dt' {\bf E}_{pump}(t')) $ where we have chosen in initial condition such that $\Av_{pump}(t)\to0$ for $t\to-\infty$. This choice of initial condition is, of course, immaterial and represents the gauge freedom of the problem. We show in the appendix that within the formalism we use in this paper\cite{sentef2015theory, freericks2009theoretical, freericks2014gauge} this choice of initial condition does not change any of our observations. 

Let us define the frequency scale associated with the pump pulse envelope $\Omega_{pump}=2\pi/T_{pump}$. We work in the limit
$ \Omega_{pump}\ll \Omega$ in which case it is appropriate to write (see Appendix~\ref{ap:PumpEnvelope} for more details)
\beq
\Av_{pump}(t)= \frac{E_0}{\Omega} e^{-\frac{t^2}{2T_{pump}^2}} \tilde{\bf E}_{\Omega}(t)
\eeq
where $\tilde{\bf E}_{\Omega}(t)$ is defined through $\frac{d}{dt} \tilde{\bf E}_{\Omega}(t)=-\Omega {\bf E}_{\Omega}(t)$. The evolution of our time-dependent system is now described through a minimal coupling of the above pump field to our Dirac Hamiltonian via $\hbar\kv\to\hbar \kv-e\Av_{pump}(t)$. Thus sidebandsthe (time-dependent) Hamiltonian we work with is as follows
\beq
H_{\kv}(t) = v_F\left[(\hbar\kv-e\Av_{pump}(t))\times \vec{\sigma} \right]\cdot \hat{z}-\mu \sigma_0
\eeq

To complete our discussion of the models we must define the probe pulse profile. For this we take the envelop function $s(t, t_{\mathcal{O}}) = e^{-\frac{(t-t_{\mathcal{O}})^2}{2T_{probe}^2}}$ where $T_{probe}$ is the width of the probe, assumed to be much shorter than the width of the pump, $T_{probe}\ll T_{pump}$, and $t_{\mathcal{O}}$ is the delay time between the pump and probe peaks. $t_{\mathcal{O}}$ is effectively the time at which we are ``viewing" the system. In the above model we have (arbitrarily) assigned $t=0$ to be the time at which the pump pulse is maximal. 

In our simulation we take experimentally relevant values for the parameters from Ref.~[\onlinecite{Wang}]. Namely, we estimate $\hbar v_F\simeq 3.6$ eV\AA, $\mu\simeq 300$meV, $\hbar\Omega\simeq$ 120meV. For convenience we define  $\omega_F=\mu/\hbar$.  We take a pump-pulse with a full width half-max (FWHM) of $250$fs ($T_{pump}\simeq 106.16$fs) and a probe-pulse with\cite{sentef2015theory} $T_{probe}=26$fs. Finally, to fully illustrate the conceptual power of our findings we take $E_0\simeq 7.5\times 10^{-3}$ V/$\AA$, slightly exaggerated from the estimates of Ref.~[\onlinecite{Wang}].

\subsection{Photocurrent}

A simplified picture of the technology involved in ARPES is to think of the experimental set-up as measuring the particle current of electrons ejected from the sample at a wave vector $\kv$, energy $\hbar\omega$ and time $t_{\mathcal{O}}$ (relative to the pump maximum time). This measurement is called the photocurrent, $I(k_x,k_y, \omega, t_{\mathcal{O}})$. Typically this quantity involves complicated momentum, orbital, and time dependent matrix elements. To develop a solid understanding for this problem we will work under the assumption that these matrix elements are the same for all orbitals, momenta and times. Under this approximation the relevant quantity to calculate is \cite{sentef2015theory, freericks2009theoretical, freericks2014gauge}
\begin{widetext}
\beqa\label{photocurrent}
&&I(k_x,k_y, \omega, t_{\mathcal{O}})=  \text{Im}\left[\int dt_1\int dt_2 s(t_1,t_{\mathcal{O}}) s(t_2,t_{\mathcal{O}})e^{i\omega (t_1-t_2)} \text{Tr}\left( G^{<}_{\kv}(t_1,t_2)\right)\right]
\eeqa
\end{widetext}
In the above $ G^{<}_{\kv}(t_1,t_2)$ is the $2\times2$ lesser Green's function matrix of the system in spin space. It is obtained by evolving the equilibrium states of the original Dirac cone from the distant past to the present.  It is defined as follows
\beq
G^{<}_{\kv\sigma\sigma'}(t,t') \equiv i\langle c^\dagger_{\kv\sigma}(t)c_{\kv\sigma'}(t')\rangle
\eeq
where $c^\dagger_{\kv\alpha}$ creates and electron with momenta $\kv$ and spin $\alpha$.

Our theory relies on knowing the solutions to the Dirac equation at all times, as these states can be used to construct the above Green's function. We therefore define the states
\beq
i\hbar \partial_t |\psi_{\kv,\alpha}(t)\rangle = H_{\kv}(t) |\psi_{\kv,\alpha}(t)\rangle
\eeq
subject to the initial condition $|\psi_{\kv,\alpha}(t\to -\infty)\rangle=|\phi_{\kv,\alpha}\rangle$ where $|\phi_{\kv,\alpha}\rangle$ are the eigenstates of the equilibrium system satisfying $h_\kv|\phi_{\kv,\alpha}\rangle=E_{\kv\alpha}|\phi_{\kv,\alpha}\rangle$, with $E_{\kv\alpha}=\alpha \hbar v_F |\kv|-\mu$ with $\alpha=\pm1$ labeling the chirality of the state. 

Once these wave functions are known the lesser Green's function of the system can be constructed (see Appendix \ref{ap:GF})
\beqa\label{eq:GLesser}
G^{<}_{\kv\sigma\sigma'}(t,t')= i\sum_{\alpha}|\psi^\sigma_{\kv,\alpha}(t)\rangle\langle \psi^{\sigma'}_{\kv,\alpha}(t')|f(E_{\kv\alpha})
\eeqa
where $f(E_{\kv\alpha})$ is a Fermi function and $|\psi^\sigma_{\kv,\alpha}(t)\rangle$ is the spin $\sigma$ component of the state $|\psi_{\kv,\alpha}(t)\rangle$.

The theory described in the rest of this paper involves determining $G^{<}_{\kv}(t,t')$ either analytically or numerically and then making use of Eq.~(\ref{photocurrent}) to estimate the results of a TR-ARPES experiment.  

\section{Results and Discussion}\label{sec:results}
Using the methodology outlined above, we present our results and interpretation of calculations relevant to TR-ARPES measurements. For the sake of clarity, we divide our discussion into two categories. First, we consider light polarized along the $x$ direction of the sample. Next, we allow for circularly polarized light. Certain limits of these two set-ups can be solved analytically and crucial insight can be gained into the distribution of states in a non-equilibrium system. We begin with linearly polarized light. 

\subsection{Linearly Polarized Light}

\begin{figure*}[]
  \setlength{\unitlength}{1mm}

   \includegraphics[scale=.5]{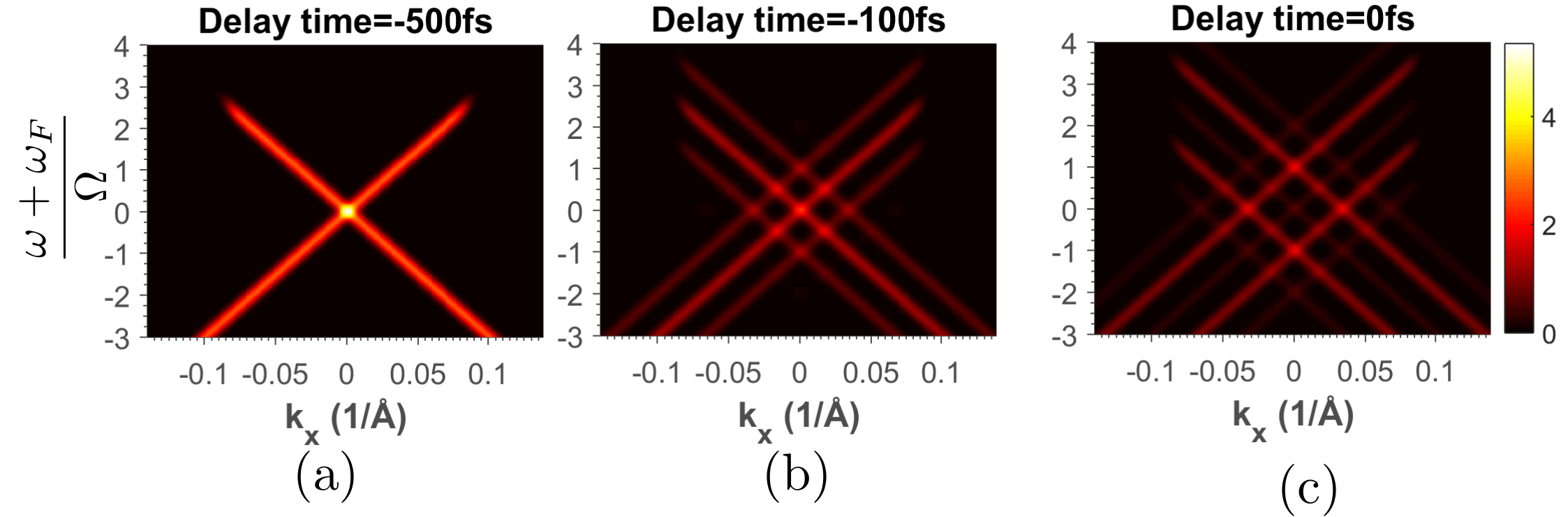}
\caption{{\small
Plot of the photocurrent $I(k_x,0,\omega,t_\mathcal{O})$ at various values of the delay time $t_\mathcal{O}$ for linearly polarized light. In the distant past we see only the Dirac cone, as the pump field starts to turn on we see copies of this cone (sidebands) begin to develop. As the field becomes full turned on the weight of the original Dirac cone is shifted into other sidebands. 
     }
     }\label{fig:ARPESlinear1}
\end{figure*}

\begin{figure*}[]
  \setlength{\unitlength}{1mm}

   \includegraphics[scale=.5]{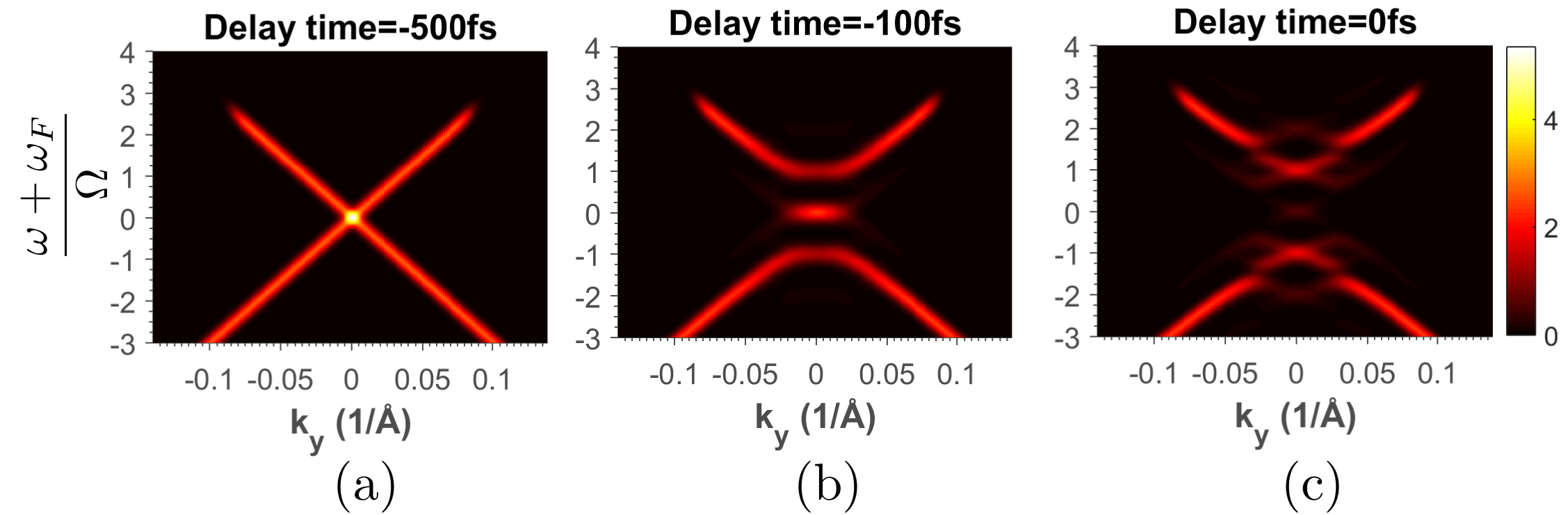}
\caption{{\small
Plot of the photocurrent $I(0,k_y,\omega,t_\mathcal{O})$ at various values of the delay time $t_\mathcal{O}$ for linearly polarized light. In the distant past we again only see the Dirac cone, as the pump field starts to turn on we see copies of this cone (sidebands) begin to develop and these copies develop avoided crossings. As the field becomes full turned on the weight of the original Dirac cone is shifted into other sidebands and these gaps become more evident.
     }
     }\label{fig:ARPESlinear2}
\end{figure*}

We consider an electric field along the $x$ direction only. In this case a closed form solution to the Dirac equation can be found along the $k_y=0$ cut of the Brillouin zone:
\beq
H_{k_x, k_y=0}(t)= v_F (\hbar k_x-eA_x(t))\sigma_y -\mu\sigma_0
\eeq
In this case the time dependent field commutes with the Hamiltonian for the chosen momenta and the wave functions can be written as
\beqa
|\psi_{k_x, 0, \alpha}(t)\rangle &=&  e^{-i( \alpha v_Fk_x-\mu/\hbar)(t-t_r)}\\ \nonumber &\times& e^{ie\alpha v_F\int_{t_r}^t dt' A_x(t')/\hbar }  |\phi_{k_x,0,\alpha}\rangle 
\eeqa
where $t_r\to-\infty$ is the ``turn-on" time for the field. This gives the Green's function 
\beqa
&&G^{<}_{k_x,0'}(t,t')= i\sum_{\alpha}e^{-i( \alpha v_Fk_x-\mu/\hbar)(t-t')}\\ \nonumber &\times& \exp\left(i\frac{e\alpha v_F}{\hbar}\int_{t'}^t dt'' A_x(t'')\right)f(E_{k_x,0\alpha}) \\ \nonumber &\times& |\phi_{k_x,0,\alpha}\rangle\langle \phi_{k_x,0,\alpha}|
\eeqa
note that the above is independent of $t_r$. We eventually need $\text{Tr}\left(G^{<}_{k_x,0'}(t,t')\right)$, where the trace is over spin degrees of freedom. This trace can be performed in any complete basis and becomes particularly simple when we choose the states $|\phi_{k_x,0,\alpha}\rangle$, which leaves
\beqa
\text{Tr}\left[G^{<}_{k_x,0'}(t,t')\right]&=& i\sum_{\alpha}e^{-i( \alpha v_Fk_x-\mu/\hbar)(t-t')} \\ \nonumber &\times& \exp\left(i\frac{e\alpha v_F}{\hbar}\int_{t'}^t dt'' A_x(t'')\right)f(E_{\kv\alpha})
\eeqa
%
Our discussion has been exact until this point.  We would now like to make an approximation to simplify the above trace.  We recall that $A_x(t)=\frac{E_0}{\Omega} e^{-t^2/2T_{pump}^2} \cos{\Omega t}$ and expand it in the limit $T_{pump}\gg 2\pi/\Omega$. Integration by parts may be used to show that to leading order in $1/T_{pump}\Omega$
\beqa
&&\int_{t'}^t dt'' A_x(t'') = \\ \nonumber && \frac{E_0}{\Omega^2} \left(e^{-t^2/2T_{pump}^2} \sin{\Omega t}-e^{-t'^2/2T_{pump}^2} \sin{\Omega t'}\right)
\eeqa
 Using the above, the identity $e^{ix\sin{\Omega t}}=\sum_{m} J_m(x)e^{im\Omega t}$ and assuming the probe pulse is much shorter than the pump pulse gives the following result for the photocurrent (for technical details see Appendix \ref{ap:Linearly})
\begin{widetext}
\beqa\label{Linearfinal}
I(k_x,0, \omega, t_{\mathcal{O}})&=&2\pi T_{probe}^2 \sum_{\alpha, m}  f(E_{k_x,0,\alpha})J_{m}^2\left(A_{eff}(t_{\mathcal{O}})\right)\exp\left[-(\omega-\alpha v_Fk_x+\frac{\mu}{\hbar}-m\Omega)^2T_{probe}^2\right]
\eeqa
\end{widetext}
where 
\beq
A_{eff}(t_{\mathcal{O}}) \simeq \frac{\int dt e^{-\frac{(t-t_{\mathcal{O}})^2}{2T_{probe}^2}}  \mathcal{A}(t)}{\int dt e^{-\frac{(t-t_{\mathcal{O}})^2}{2T_{probe}^2}} }
\eeq
 with $ \mathcal{A}(t)= \frac{e E_0v_F}{\hbar \Omega^2}e^{-t^2/2T_{pump}^2}$. The above formula is our main analytic result for this part of the paper. It provides a nice picture of the side-band splitting that occurs in the presence of a periodic field. Owing to the nature of the applied field, which commutes with the Hamiltonian, none of the original bands are dressed. The exponent describes peaks not just at energy eigenvalues $v_F\hbar k_x-\mu$, but also at integer values of $\hbar\Omega$ above and below this value. This indicates that there are copies of the original band structure at multiples of $\hbar\Omega$ above and below the original pattern. 

The Bessel function pre-factor gives the weights of these side-band peaks. These weights depend 
on the probe time due to the time dependence of the driven system.
The physical picture which emerges here is as follows.  While the periodic part of the pump pulse is responsible for the existence of side bands and their dispersion, their relative contribution to the photocurrent is given by the above Bessel functions.  Therefore, the ration of the pump envelope to its frequency determine the side band weights at any given time.  In addition, it is evident from Eq.~\ref{Linearfinal} that the probe pulse determines the time resolution, as the effective gauge field is a weighted average of the pump over the probe duration.

Owing to the simple structure at $k_y=0$ there is no interference/avoided crossing of sidebands.  Thus in the limit of a wide pump pulse the system is split into sidebands and the population of these side bands is given by the instantaneous weighted average of the pump envelope function. 

We now turn to numerics in order to test the validity of our analytic results and to extend our analysis  to finite $k_y$. For this we integrate the Dirac equation numerically. We begin by fixing $k_y=0$ and comparing our analytic treatment to exact numerics. Fig.~\ref{fig:linearcompare} shows $I(k_x,0,\omega,  t_{\mathcal{O}})$ for $k_x=0$ and $k_x=0.05\AA^{-1}$ for several values of $ t_{\mathcal{O}}$. As can be seen in the figure, there is excellent agreement between our approximate formula above and the numerics. Fig.~\ref{fig:linearcompare} also nicely illustrates the side-band interpretation discussed above. We see that all of the spectral weight associated with the original peaks in the distant past (before the pump pulse hits the system) gets redistributed into sidebands separated by $\hbar\Omega$. 

Next we move on to present results going beyond the scope of the analytic results. We plot $I(k_x,0,\omega,  t_{\mathcal{O}})$ and $I(0,k_y,\omega,  t_{\mathcal{O}})$ in Figs.~\ref{fig:ARPESlinear1} and \ref{fig:ARPESlinear2}. First, the results for $I(k_x,0,\omega,  t_{\mathcal{O}})$ (within the purview of the analytic approach above) nicely confirm the intuition developed above; we see {\it no} renormalization of the energy bands and a simple development of of sidebands. These sidebands are evident by the copies of the Dirac cone seen in the above plots. Second, $I(0,k_y,\omega,  t_{\mathcal{O}})$ goes beyond our analytic approach above. We see a twofold effect as the pump-pulse hits the system. The primary effect is a splitting of the system into side bands. Secondary, we see that the light renormalizes the side-band structure opening gaps in energies where level crossing occurs at equilibrium.

\subsection{Circularly Polarized Light}
\begin{figure*}[]
  \setlength{\unitlength}{1mm}

   \includegraphics[scale=.55]{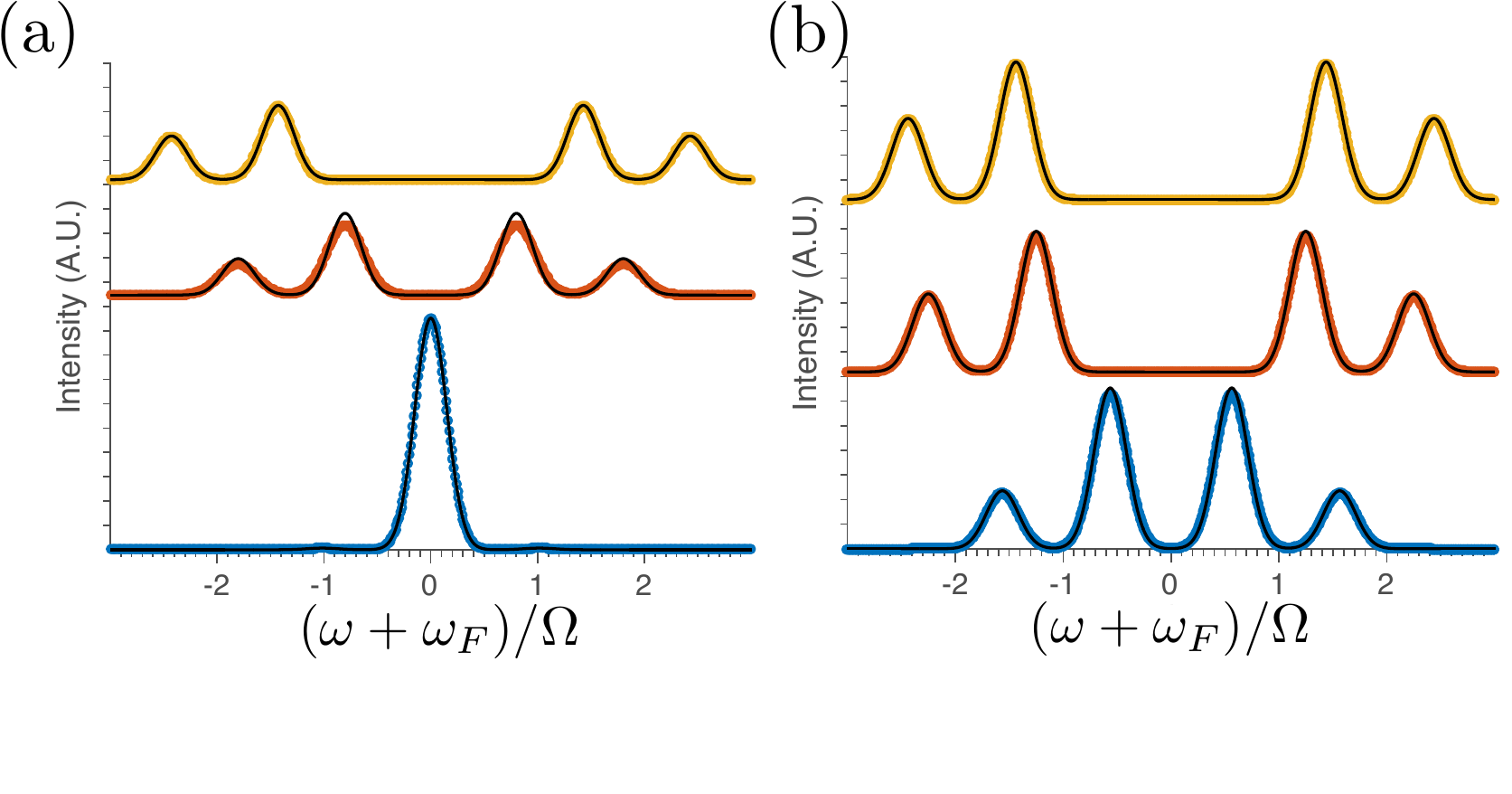}
\caption{{\small Comparison of numerical results found by integrating the time dependent Dirac equation and the analytic approximation in Eq.~(\ref{circlefinal}) for $I(0,0,\omega,t_\mathcal{O})$. The left shows $I(0,0,\omega,t_\mathcal{O})$ for different delay times $t_\mathcal{O}$ while the right shows $I(0,0,\omega,t_\mathcal{O})$  also for different delay times but this time with a pulse FWHM of $500$fs instead of $250$fs. In these plots the bottom plot is for $t_\mathcal{O}=-500$fs, the middle for $t_\mathcal{O}=-100$fs and the top is for $t_\mathcal{O}=0$fs. In all plots the solid line is the approximation in  Eq.~(\ref{circlefinal}) while the circles are numerical results. On the top row we see reasonable agreement between numerics and our approximation for all three delay times. When we turn up the pulse width, which effectively makes the "turn-on" time slower, we see that the agreement becomes excellent. 
     }
     }\label{fig:DiracCircular}
\end{figure*}

\begin{figure*}[]
  \setlength{\unitlength}{1mm}

   \includegraphics[scale=.5]{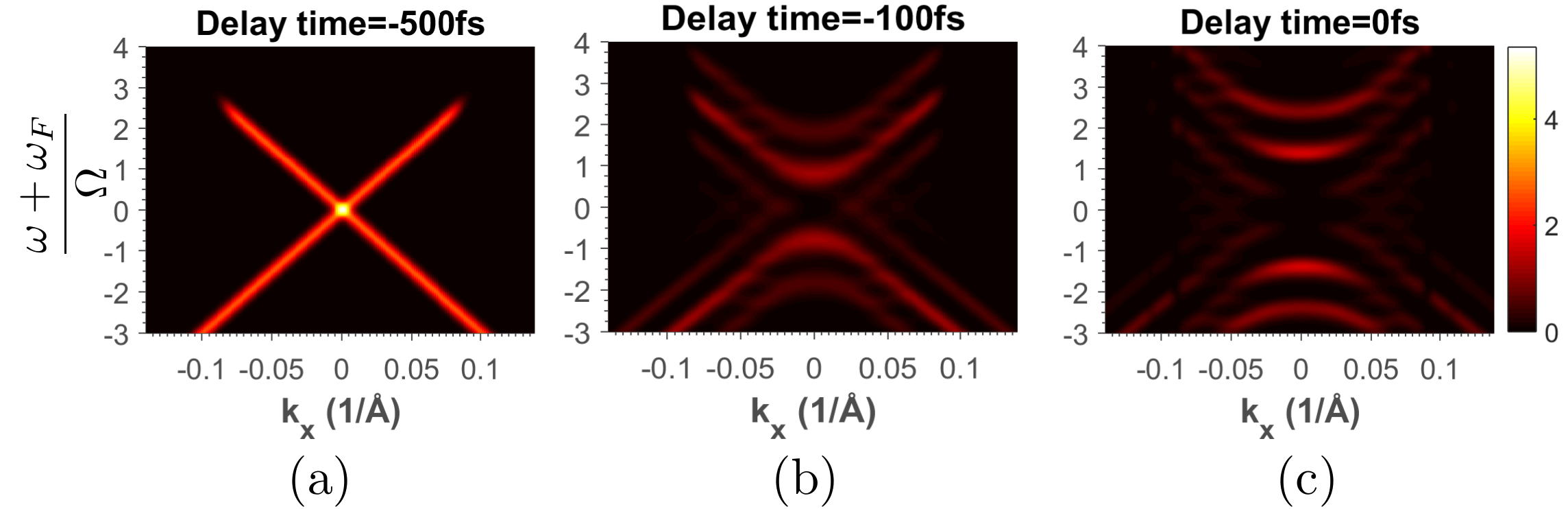}
\caption{{\small
Plot of the photocurrent $I(k_x,0,\omega,t_\mathcal{O})$ at various values of the delay time $t_\mathcal{O}$ for circularly polarized light. In the distant past we see only the Dirac cone, as the pump field starts to turn on we see copies of this cone (sidebands) begin to develop and the Dirac cone becomes gapped out. At $t_\mathcal{O}=0$ we can plainly see only two sidebands through the cut $k_x=0$. 
     }
     }\label{fig:ARPEScircle1}
\end{figure*}
beyond
%

We now shift our focus to the more involved problem of circularly polarized light. Circularly polarized light makes even the $k_y=0$ cut along momentum space intractable analytically. We can, however, make progress right at the equilibrium Dirac point $k_x=k_y=0$, the $\Gamma$-point. Here we have
\beq
H_{\Gamma}(t)=-  \hbar \Omega \mathcal{A}(t) \left[\cos{\Omega t} \sigma_y-\sin{\Omega t}\sigma_x\right]-\mu \sigma_0
\eeq
The above can be written as
\beq
H_{\Gamma}(t)=-  \hbar \Omega \mathcal{A}(t)  e^{-i\Omega t\sigma_z/2}\sigma_ye^{i\Omega t\sigma_z/2}-\mu \sigma_0
\eeq
\\
To solve for the evolution under this Hamiltonian we transform to a rotating frame by letting $|\psi_{\Gamma \alpha}(t)\rangle = e^{i\mu(t-t_r)/\hbar} e^{i\Omega t \sigma_z/2} |\hat{\psi}_{\Gamma \alpha}(t)\rangle $. Our equation of motion for the wave function then reads
\beq\label{eq:eom}
i\hbar \partial_t  |\hat{\psi}_{\Gamma \alpha}(t)\rangle =\left( - \hbar \Omega\mathcal{A}(t) \sigma_y+\frac{\hbar\Omega}{2} \sigma_z  \right) |\hat{\psi}_{\Gamma \alpha}(t)\rangle
\eeq
Our purpose in finding $|\psi_{\Gamma \alpha}(t)\rangle$ is to build the Green's function $G^<_{\Gamma}(t,t')$ and ultimately convolve this Green's function with the probe pulse envelope. Therefore, a good first approximation would be to find the wave function in the vicinity of $t_{\mathcal{O}}$, the peak time of the probe-pulse. We therefore make the somewhat crude approximation $\mathcal{A}(t)\to A_{eff}(t_{\mathcal{O}})$ in the above equation of motion, Eq.~\ref{eq:eom}. This yields an effective (rotating frame) Hamiltonian which is time independent.  The above equation of motion can be therefore solved to give: 
\beqa
&&|\psi_{\Gamma \alpha}(t)\rangle = e^{i\mu(t-t_r)/\hbar} e^{i\Omega t \sigma_z/2}\\ \nonumber &\times& e^{-iH_{eff}(t-t_r)/\hbar}  |\hat \psi_{\Gamma \alpha}(t_r)\rangle
\eeqa
where
\beq 
H_{eff}=- \hbar \Omega A_{eff}(t_{\mathcal{O}}) \sigma_y+\frac{\hbar\Omega}{2} \sigma_z
\eeq
is an effective, time-independent, Hamiltonian in the vicinity of $t_{\mathcal{O}}$.  $|\hat \psi_{\Gamma \alpha}(t_r)\rangle$  is the rotating frame wave function in the distant past.  By requiring that the wavefunction $|\psi_{\Gamma \alpha}(t)\rangle$ at $t\to t_r$ be a simple spinor (whose direction can be arbitrarily chosen due to the degeneracy at the Dirac point) we find 
\begin{equation}
 |\hat \psi_{\Gamma \alpha}(t_r)\rangle = e^{-i\Omega t_r \sigma_z/2} |\phi_{\Gamma \alpha}\rangle 
\end{equation}
where $\{|\phi_{\Gamma \alpha}\rangle \}$ are the eigenstates at the Dirac point in the distant past. Using the above, noting that the eigenvalues of  $H_{eff}$ are $\pm E_{eff}(t_{\mathcal{O}})= \pm \sqrt{(\hbar v_FA_{eff}(t_{\mathcal{O}}))^2+(\frac{\hbar\Omega}{2})^2}$, and performing some additional manipulations which are left for Appendix \ref{ap:Circularly} we arrive at the following approximation for the photocurrent
\begin{widetext}
 \beqa
&&I(0,0, \omega, t_{\mathcal{O}})=2\pi T_{probe}^2\text{Im}\left[ \sum_{\alpha\beta, s,s'} f(\epsilon_{\Gamma\alpha})e^{-i(s-s')E_{eff}(t_{\mathcal{O}})t_r/\hbar} A^{s}_{\alpha,\beta}(A^{s'}_{\alpha,\beta})^*\right. \\\nonumber &\times& \left.\exp\left[-(\omega+\mu/\hbar-\beta\Omega/2-sE_{eff}(t_{\mathcal{O}})/\hbar)^2T_{probe}^2/2\right]   \exp\left[-(\omega+\mu/\hbar-\beta\Omega/2-s'E_{eff}(t_{\mathcal{O}})/\hbar)^2T_{probe}^2/2\right]\right]
\eeqa
\end{widetext}
 where $\alpha,\beta, s, s'$ all run over $\pm1$ and $A^{s}_{\alpha,\beta}= \left( \delta_{\alpha,\beta}-s \hat{a} \cdot    \left \langle \phi_{\beta}\left |\vec{\sigma} \right |\phi_{\alpha}\right\rangle\right)/2$, $|\phi_{+}\rangle =(1,0)^T$, $|\phi_{-}\rangle =(0,1)^T$  and ${\bf a} = a  \hat{a}=  - \hbar \Omega A_{eff}(t_{\mathcal{O}}) \hat{y}+\frac{\hbar\Omega}{2} \hat{z}$. 

 We now note that $E_{eff}(t_{\mathcal{O}})=\sqrt{(\hbar v_FA_{eff}(t_{\mathcal{O}}))^2+(\frac{\hbar\Omega}{2})^2}\ge \hbar\Omega/2 \gg1/T_{probe}$. The gap betweem states is $2E_{eff}(t_{\mathcal{O}})/\hbar$, which is consistent with the gap found in [\onlinecite{fregoso}] in a purely Floquet system ($T_{probe}\to\infty$). Importantly, the distance separating the peaks in the Gassians above is much larger than the width of the peaks. We therefore discard terms where $s\ne s'$. Further, we note that the eigenvalues in the distant past $\epsilon_{\Gamma\alpha}=-\mu$ are independent of $\alpha$ (as we're at the Dirac point). These two observations along with some additional straightforward, but tedious, algebra lead to the simplified result
\begin{widetext} 
  \beqa\label{circlefinal}
I(0,0, \omega, t_{\mathcal{O}})&=&2\pi T_{probe}^2  f(-\mu) \sum_{\beta, s} \left(\frac{1-s\beta \hat{a}_z}{2}\right)\exp\left[-(\omega+\mu/\hbar-\beta\Omega/2-sE_{eff}(t_{\mathcal{O}})/\hbar)^2T_{probe}^2\right]
\eeqa
\end{widetext}
Examining the above shows that the ARPES spectrum from the $\Gamma$ point shows the following features at energies $E$ (measured from $\mu$) with weights $P$
\beqa
&&E_1= -E_{eff}(t_{\mathcal{O}})+\hbar\Omega/2 \ \ \ \  P_1= \left(\frac{1+\hat{a}_z}{2}\right) \\ \nonumber
&&E_2= E_{eff}(t_{\mathcal{O}})-\hbar\Omega/2 \ \ \ \  P_2= \left(\frac{1+\hat{a}_z}{2}\right) \\ \nonumber
&&E_3= -E_{eff}(t_{\mathcal{O}})-\hbar\Omega/2 \ \ \ \  P_3= \left(\frac{1-\hat{a}_z}{2}\right) \\ \nonumber
&&E_4= E_{eff}(t_{\mathcal{O}})+\hbar\Omega/2 \ \ \ \  P_4= \left(\frac{1-\hat{a}_z}{2}\right) \\ \nonumber
\eeqa
where $\hat{a}_z=\frac{\hbar\Omega}{ 2E_{eff}(t_{\mathcal{O}})}$. It is obvious from the above the there's no additional spectral weight in any other energy.  When the amplitude $A(t)$ is shut off $\hat{a}_z\to1$ and we see $E_1,E_2 \to 0$ with weights going to unity. At the same time $E_3, E_4\to \pm \hbar \Omega$, albeit with zero weight. 

Our interpretation of the above is as follows. As the pump probe is turned on, the original two-fold degeneracy at the Dirac point is lifted and a gap is opened up with width
\beq
G(t_{\mathcal{O}}) = \sqrt{(2\hbar v_FA_{eff}(t_{\mathcal{O}}))^2+(\hbar\Omega)^2}-\hbar\Omega
\eeq
the weight of these states is $ \left(\frac{1+\hat{a}_z}{2}\right)$ which decreases with field strength. The peaks at the other two energies correspond to single sidebands of the states $E_1$ and $E_2$. Put another way,  $E_3=E_1-\Omega\hbar$ while $E_4=E_2+\hbar\Omega$. The weights of these side bands increase with field strength. Interestingly, unlike our treatment of the linearly polarized light, there is no statistical weight given to any other side bands, all of the spectral weight is found within two sidebands. Note that the same approximations were made in both cases.  {We can trace this phenomenon back to the transformation we made to the rotating frame.  While in general, this transformation leaves the Hamiltonian time dependent, here it does not since we have used an effective field strength. In the rotating frame we find two solutions to our Hamiltonian and transforming back to the original frame can split each one of these into two side band due to the dimensionality of the transformation operator.  It is interesting to note that the same behavior was found by Dehghani {\it et al.} in Ref.~[\onlinecite{Dehghani1}].}

With the above analytic analysis let us move on to numerical methods in an effort to validate the above description and further explore momenta where an approximate solution is not tractable. We do this with the side-band language discussed above in mind. 

%

We begin with a simulation at the Gamma point. Fig.~\ref{fig:DiracCircular} shows $I(0,0, \omega, t_{\mathcal{O}})$ as a function of $\omega$ for various different values of $ t_{\mathcal{O}}$. Both our approximate analytic expression as well as our numerics are displayed in this plot. We see that the approximation provided above is in good agreement with the numerics with respect to both the size of the gap and the position of the sidebands, it also shows that this approximation becomes better when the width of the pump-pulse gets larger.  

Before continuing we would like to highlight the fact that the discussion here appears to be more general than the Dirac model we have used. To check that our conclusions are not simply a coincidence of this model we have gone beyond the Dirac cone model by including higher order corrections\cite{FuWarp} and also by studying a lattice model for TIs. Our conclusions of only two side-bands do not change. In order to keep the discussion of the main text simple we have included the details of calculations on these models in the appendix. 

Let us now move on to explore a wider range of momentum using our numerical protocol. Fig.~\ref{fig:ARPEScircle1}  shows the time-evolution of the ARPES spectrum for a cut such that $k_y=0$. A cut along $k_x=0$ looks very similar and such plots would not add to the present discussion. In the figure we see effects common to all results in this work. As the field strength is turned on the original Dirac cone is copied into sidebands, each of which is populated only with a certain weight. States in these sidebands then hybridize with each other leading to gaps. 
%
Most notable, our analytic result for the $\Gamma$ point is verified at the center of the momentum cut.

\subsection{Timescales}

\begin{figure}[]
  \setlength{\unitlength}{1mm}

   \includegraphics[scale=.45]{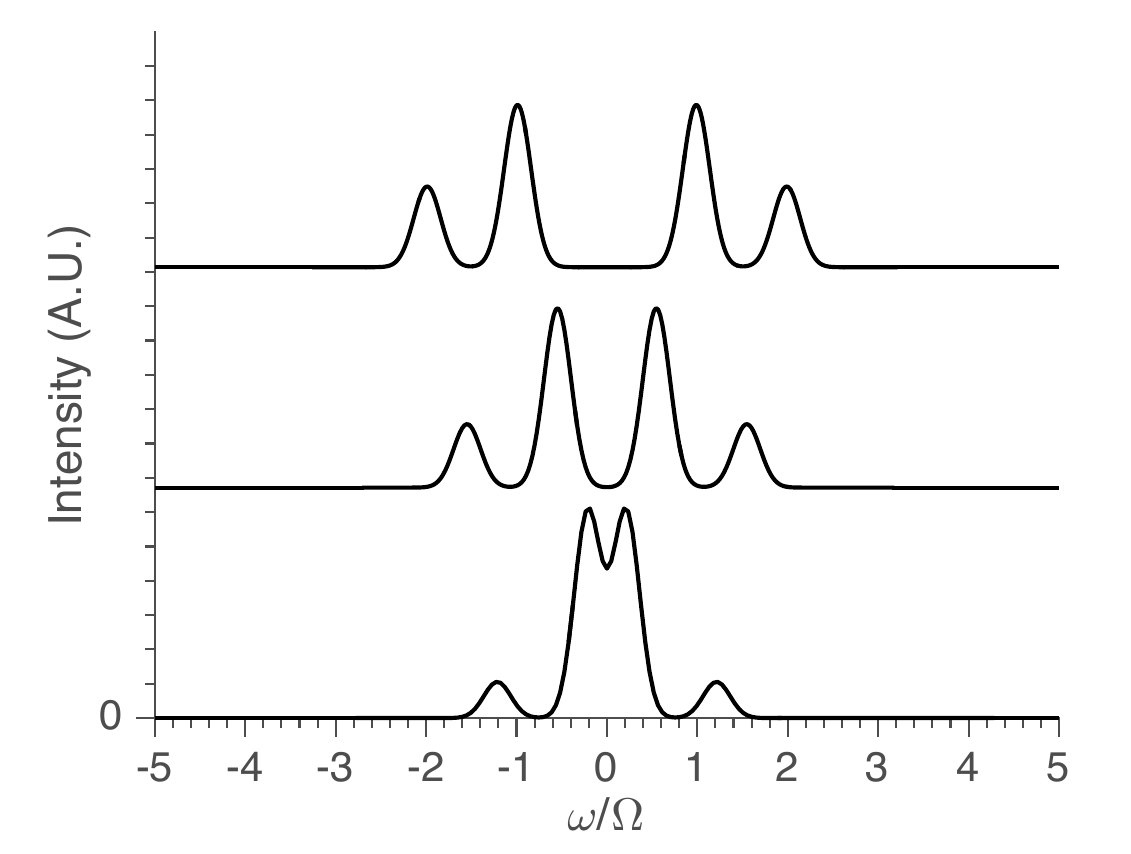} \\
   \includegraphics[scale=.45]{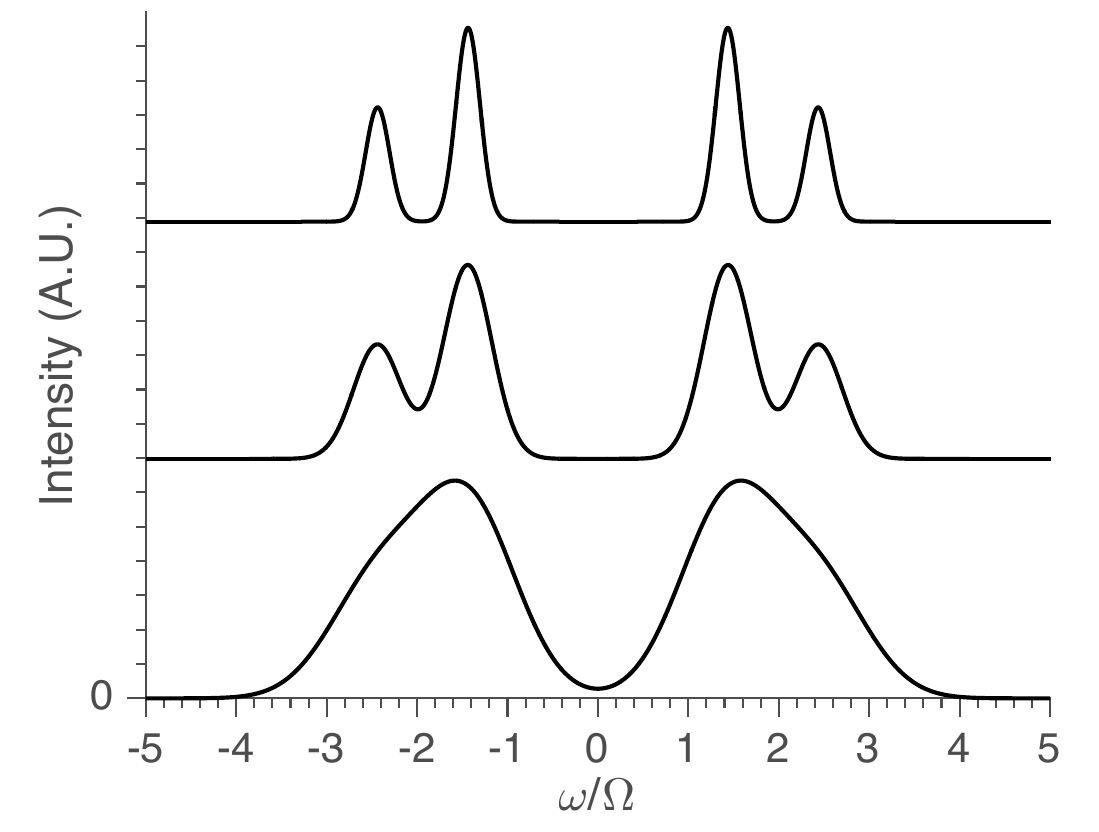}
\caption{{\small Plot of $I(0,0,\omega,0)$ for circularly polarized light at varying $T_{pump}$ values, top, and at varying $T_{probe}$ values on the bottom. In the top plot we have plotted $I(0,0,\omega,t_\mathcal{O})$  for a pump pulse with a FWHM of $\pi/\Omega$ (top), $2\pi/\Omega$ (middle), and $4\pi/\Omega$ (bottom). Similarly in the bottom plot we have plotted $I(0,0,\omega,t_\mathcal{O})$  for a probe pulse with a FWHM of $\pi/\Omega$ (top), $2\pi/\Omega$ (middle), and $4\pi/\Omega$ (bottom). In the top plot we see the development of sidebands despite the pump envelope only containing a minimum number of oscillations. In the bottom plot we see that the probe pulsewidth sets our ability to resolve the side-band peaks. 
     }
     }\label{fig:timescales}
\end{figure}

We close this work with a short discussion on the timescales required to see the side-band physics that we have discusses here. In the majority of this paper we have focused on a hirarchy $T_{pump}\gg T_{probe} \gg 2\pi/\Omega$. Focusing on these timescales was not only relevant from an experimental viewpoint\cite{Wang} but also aiding in our derivation of approximate analytic results. Here we will briefly explore what happens when these conditions are relaxed. First, we have explored the effect of varying $T_{pump}$ on the development of side-bands. We have found that varying $T_{pump}$ down to even half of $T_\Omega=2\pi/\Omega$ one can {\em still} see the development of side-bands. Results of this can be seen in the top plot of Fig. \ref{fig:timescales}. We see that the major effect of varying $T_{pump}$ is that the {\em effective} field strength that the system sees is decreased (visible by noting the smaller gap and side-band amplitudes). We can understand this heuristically as the system having less time to {\em see} the field while the envelope is at large values. Second, we have studied how changing $T_{probe}$ can change our observations. Not surprisingly, $1/T_{probe}$ sets our energy resolution in the photocurrent. A very narrow probe width leads to very broadened side-band peaks that can overlap with each other. For sharp side-band peaks the width of the probe field should be made as large as possible. The intuition behind this appears to be that the probe field needs to observe the system for at least a few periods, $T_{\Omega}$, to properly observe the sidebands. Our results are summarized in the bottom panel of Fig. \ref{fig:timescales}. 

\section{Conclusions}

We have provided both simple analytic results and rigorous numerical simulations of TR-ARPES in a Dirac system. Our results show that the time-evolution of an ARPES spectrum can be understood using the language of probabilistic occupation of sidebands in a time periodic system\cite{Farrelltop, Farrelltriv}. 
The population of these side bands depends non-linearly on the envelope of the pump field.  As expected, when the pump is turned off gradually the side bands are not populated while the original band is populated with probability 1.  The shape of the probe pulse determines the time resolution and therefore, when the probe is sharp the TR-ARPES photocurrent follows the time evolution of the side band population.
Our results are in qualitative agreement with those of the experimental work in Ref.~[\onlinecite{Wang}].

%
Our work also highlights the fact that not all side bands are equally important.  We showed that despite the repeated folding of the Dirac cone into the Floquet zone only a few side bands, which are displaced by a few $\hbar\Omega$s from the equilibrium energy, contribute to the time resolved ARPES signal.  This point pertains not only to the results of Ref.~[\onlinecite{Wang}] but also to other measurements on Floquet topological systems such as transport\cite{Farrelltriv,Farrelltop}.

Finally, we have explored an interesting interplay between a Dirac point and circularly polarized light. Our results suggest that under the application of circularly polarized light the Dirac point is gapped and copied into two sidebands {\em only}. This is in contrast to other systems, and other points in the Brillouin zone, where many sidebands can be seen. 

\section{Acknowledgments} 
The authors are thankful for useful discussions with Nuh Gedik. Financial support for this work was provided by the NSERC and FQRNT (TPB) and the Vanier Canada Graduate Scholarship (AF). Numerical calculations for this work were performed using McGill HPC supercomputing resources.
\bibliographystyle{apsrev}
\bibliography{Floquet}

\begin{thebibliography}{47}
\expandafter\ifx\csname natexlab\endcsname\relax\def\natexlab#1{#1}\fi
\expandafter\ifx\csname bibnamefont\endcsname\relax
  \def\bibnamefont#1{#1}\fi
\expandafter\ifx\csname bibfnamefont\endcsname\relax
  \def\bibfnamefont#1{#1}\fi
\expandafter\ifx\csname citenamefont\endcsname\relax
  \def\citenamefont#1{#1}\fi
\expandafter\ifx\csname url\endcsname\relax
  \def\url#1{\texttt{#1}}\fi
\expandafter\ifx\csname urlprefix\endcsname\relax\def\urlprefix{URL }\fi
\providecommand{\bibinfo}[2]{#2}
\providecommand{\eprint}[2][]{\url{#2}}

\bibitem[{\citenamefont{Farrell and Pereg-Barnea}(2015)}]{Farrelltop}
\bibinfo{author}{\bibfnamefont{A.}~\bibnamefont{Farrell}} \bibnamefont{and}
  \bibinfo{author}{\bibfnamefont{T.}~\bibnamefont{Pereg-Barnea}},
  \bibinfo{journal}{Phys. Rev. Lett.} \textbf{\bibinfo{volume}{115}},
  \bibinfo{pages}{106403} (\bibinfo{year}{2015}),
  \urlprefix\url{http://link.aps.org/doi/10.1103/PhysRevLett.115.106403}.

\bibitem[{\citenamefont{Farrell and Pereg-Barnea}(2016)}]{Farrelltriv}
\bibinfo{author}{\bibfnamefont{A.}~\bibnamefont{Farrell}} \bibnamefont{and}
  \bibinfo{author}{\bibfnamefont{T.}~\bibnamefont{Pereg-Barnea}},
  \bibinfo{journal}{Phys. Rev. B} \textbf{\bibinfo{volume}{93}},
  \bibinfo{pages}{045121} (\bibinfo{year}{2016}).

\bibitem[{\citenamefont{Wang et~al.}(2013)\citenamefont{Wang, Steinberg,
  Jarillo-Herrero, and Gedik}}]{Wang}
\bibinfo{author}{\bibfnamefont{Y.~H.} \bibnamefont{Wang}},
  \bibinfo{author}{\bibfnamefont{H.}~\bibnamefont{Steinberg}},
  \bibinfo{author}{\bibfnamefont{P.}~\bibnamefont{Jarillo-Herrero}},
  \bibnamefont{and} \bibinfo{author}{\bibfnamefont{N.}~\bibnamefont{Gedik}},
  \bibinfo{journal}{Science} \textbf{\bibinfo{volume}{342}},
  \bibinfo{pages}{453} (\bibinfo{year}{2013}).

\bibitem[{\citenamefont{Bernevig et~al.}(2006)\citenamefont{Bernevig, Hughes,
  and Zhang}}]{Bernevig}
\bibinfo{author}{\bibfnamefont{B.~A.} \bibnamefont{Bernevig}},
  \bibinfo{author}{\bibfnamefont{T.~L.} \bibnamefont{Hughes}},
  \bibnamefont{and} \bibinfo{author}{\bibfnamefont{S.-C.} \bibnamefont{Zhang}},
  \bibinfo{journal}{Science} \textbf{\bibinfo{volume}{314}},
  \bibinfo{pages}{1757} (\bibinfo{year}{2006}).

\bibitem[{\citenamefont{Moore and Balents}(2007)}]{moore}
\bibinfo{author}{\bibfnamefont{J.~E.} \bibnamefont{Moore}} \bibnamefont{and}
  \bibinfo{author}{\bibfnamefont{L.}~\bibnamefont{Balents}},
  \bibinfo{journal}{Physical Review B} \textbf{\bibinfo{volume}{75}},
  \bibinfo{pages}{121306} (\bibinfo{year}{2007}).

\bibitem[{\citenamefont{Fu et~al.}(2007)\citenamefont{Fu, Kane, and Mele}}]{fu}
\bibinfo{author}{\bibfnamefont{L.}~\bibnamefont{Fu}},
  \bibinfo{author}{\bibfnamefont{C.~L.} \bibnamefont{Kane}}, \bibnamefont{and}
  \bibinfo{author}{\bibfnamefont{E.~J.} \bibnamefont{Mele}},
  \bibinfo{journal}{Physical Review Letters} \textbf{\bibinfo{volume}{98}},
  \bibinfo{pages}{106803} (\bibinfo{year}{2007}).

\bibitem[{\citenamefont{Hsieh et~al.}(2008)\citenamefont{Hsieh, Qian, Wray,
  Xia, Hor, Cava, and Hasan}}]{hsieh}
\bibinfo{author}{\bibfnamefont{D.}~\bibnamefont{Hsieh}},
  \bibinfo{author}{\bibfnamefont{D.}~\bibnamefont{Qian}},
  \bibinfo{author}{\bibfnamefont{L.}~\bibnamefont{Wray}},
  \bibinfo{author}{\bibfnamefont{Y.}~\bibnamefont{Xia}},
  \bibinfo{author}{\bibfnamefont{Y.~S.} \bibnamefont{Hor}},
  \bibinfo{author}{\bibfnamefont{R.}~\bibnamefont{Cava}}, \bibnamefont{and}
  \bibinfo{author}{\bibfnamefont{M.~Z.} \bibnamefont{Hasan}},
  \bibinfo{journal}{Nature} \textbf{\bibinfo{volume}{452}},
  \bibinfo{pages}{970} (\bibinfo{year}{2008}).

\bibitem[{\citenamefont{K\"onig et~al.}(2007)\citenamefont{K\"onig, Wiedmann,
  Br�ne, Roth, Buhmann, Molenkamp, Qi, and Zhang}}]{konig}
\bibinfo{author}{\bibfnamefont{M.}~\bibnamefont{K\"onig}},
  \bibinfo{author}{\bibfnamefont{S.}~\bibnamefont{Wiedmann}},
  \bibinfo{author}{\bibfnamefont{C.}~\bibnamefont{Br�ne}},
  \bibinfo{author}{\bibfnamefont{A.}~\bibnamefont{Roth}},
  \bibinfo{author}{\bibfnamefont{H.}~\bibnamefont{Buhmann}},
  \bibinfo{author}{\bibfnamefont{L.~W.} \bibnamefont{Molenkamp}},
  \bibinfo{author}{\bibfnamefont{X.-L.} \bibnamefont{Qi}}, \bibnamefont{and}
  \bibinfo{author}{\bibfnamefont{S.-C.} \bibnamefont{Zhang}},
  \bibinfo{journal}{Science} \textbf{\bibinfo{volume}{318}},
  \bibinfo{pages}{766} (\bibinfo{year}{2007}).

\bibitem[{\citenamefont{Roth et~al.}(2009)\citenamefont{Roth, Br\"une, Buhmann,
  Molenkamp, Maciejko, Qi, and Zhang}}]{Roth}
\bibinfo{author}{\bibfnamefont{A.}~\bibnamefont{Roth}},
  \bibinfo{author}{\bibfnamefont{C.}~\bibnamefont{Br\"une}},
  \bibinfo{author}{\bibfnamefont{H.}~\bibnamefont{Buhmann}},
  \bibinfo{author}{\bibfnamefont{L.~W.} \bibnamefont{Molenkamp}},
  \bibinfo{author}{\bibfnamefont{J.}~\bibnamefont{Maciejko}},
  \bibinfo{author}{\bibfnamefont{X.-L.} \bibnamefont{Qi}}, \bibnamefont{and}
  \bibinfo{author}{\bibfnamefont{S.-C.} \bibnamefont{Zhang}},
  \bibinfo{journal}{Science} \textbf{\bibinfo{volume}{325}},
  \bibinfo{pages}{294} (\bibinfo{year}{2009}).

\bibitem[{\citenamefont{Li et~al.}(2009)\citenamefont{Li, Chu, Jain, and
  Shen}}]{li09}
\bibinfo{author}{\bibfnamefont{J.}~\bibnamefont{Li}},
  \bibinfo{author}{\bibfnamefont{R.-L.} \bibnamefont{Chu}},
  \bibinfo{author}{\bibfnamefont{J.~K.} \bibnamefont{Jain}}, \bibnamefont{and}
  \bibinfo{author}{\bibfnamefont{S.-Q.} \bibnamefont{Shen}},
  \bibinfo{journal}{Phys. Rev. Lett.} \textbf{\bibinfo{volume}{102}},
  \bibinfo{pages}{136806} (\bibinfo{year}{2009}).

\bibitem[{\citenamefont{Borchmann et~al.}(2016)\citenamefont{Borchmann,
  Farrell, and Pereg-Barnea}}]{borchmann}
\bibinfo{author}{\bibfnamefont{J.}~\bibnamefont{Borchmann}},
  \bibinfo{author}{\bibfnamefont{A.}~\bibnamefont{Farrell}}, \bibnamefont{and}
  \bibinfo{author}{\bibfnamefont{T.}~\bibnamefont{Pereg-Barnea}},
  \bibinfo{journal}{Phys. Rev. B} \textbf{\bibinfo{volume}{93}},
  \bibinfo{pages}{125133} (\bibinfo{year}{2016}),
  \urlprefix\url{http://link.aps.org/doi/10.1103/PhysRevB.93.125133}.

\bibitem[{\citenamefont{Lindner et~al.}(2011)\citenamefont{Lindner, Refael, and
  Galitski}}]{Lindner}
\bibinfo{author}{\bibfnamefont{N.~H.} \bibnamefont{Lindner}},
  \bibinfo{author}{\bibfnamefont{G.}~\bibnamefont{Refael}}, \bibnamefont{and}
  \bibinfo{author}{\bibfnamefont{V.}~\bibnamefont{Galitski}},
  \bibinfo{journal}{Nat Phys} \textbf{\bibinfo{volume}{7}},
  \bibinfo{pages}{490} (\bibinfo{year}{2011}).

\bibitem[{\citenamefont{Gu et~al.}(2011)\citenamefont{Gu, Fertig, Arovas, and
  Auerbach}}]{Gu}
\bibinfo{author}{\bibfnamefont{Z.}~\bibnamefont{Gu}},
  \bibinfo{author}{\bibfnamefont{H.~A.} \bibnamefont{Fertig}},
  \bibinfo{author}{\bibfnamefont{D.~P.} \bibnamefont{Arovas}},
  \bibnamefont{and} \bibinfo{author}{\bibfnamefont{A.}~\bibnamefont{Auerbach}},
  \bibinfo{journal}{Phys. Rev. Lett.} \textbf{\bibinfo{volume}{107}},
  \bibinfo{pages}{216601} (\bibinfo{year}{2011}).

\bibitem[{\citenamefont{Oka and Aoki}(2009)}]{Oka}
\bibinfo{author}{\bibfnamefont{T.}~\bibnamefont{Oka}} \bibnamefont{and}
  \bibinfo{author}{\bibfnamefont{H.}~\bibnamefont{Aoki}},
  \bibinfo{journal}{Phys. Rev. B} \textbf{\bibinfo{volume}{79}},
  \bibinfo{pages}{081406} (\bibinfo{year}{2009}).

\bibitem[{\citenamefont{Usaj et~al.}(2014)\citenamefont{Usaj, Perez-Piskunow,
  Foa~Torres, and Balseiro}}]{Usaj}
\bibinfo{author}{\bibfnamefont{G.}~\bibnamefont{Usaj}},
  \bibinfo{author}{\bibfnamefont{P.~M.} \bibnamefont{Perez-Piskunow}},
  \bibinfo{author}{\bibfnamefont{L.~E.~F.} \bibnamefont{Foa~Torres}},
  \bibnamefont{and} \bibinfo{author}{\bibfnamefont{C.~A.}
  \bibnamefont{Balseiro}}, \bibinfo{journal}{Phys. Rev. B}
  \textbf{\bibinfo{volume}{90}}, \bibinfo{pages}{115423}
  (\bibinfo{year}{2014}).

\bibitem[{\citenamefont{Calvo et~al.}(2011)\citenamefont{Calvo, Pastawski,
  Roche, and Torres}}]{Calvo}
\bibinfo{author}{\bibfnamefont{H.~L.} \bibnamefont{Calvo}},
  \bibinfo{author}{\bibfnamefont{H.~M.} \bibnamefont{Pastawski}},
  \bibinfo{author}{\bibfnamefont{S.}~\bibnamefont{Roche}}, \bibnamefont{and}
  \bibinfo{author}{\bibfnamefont{L.~E. F.~F.} \bibnamefont{Torres}},
  \bibinfo{journal}{Applied Physics Letters} \textbf{\bibinfo{volume}{98}}
  (\bibinfo{year}{2011}).

\bibitem[{\citenamefont{Foa~Torres et~al.}(2014)\citenamefont{Foa~Torres,
  Perez-Piskunow, Balseiro, and Usaj}}]{Torres}
\bibinfo{author}{\bibfnamefont{L.}~\bibnamefont{Foa~Torres}},
  \bibinfo{author}{\bibfnamefont{P.}~\bibnamefont{Perez-Piskunow}},
  \bibinfo{author}{\bibfnamefont{C.}~\bibnamefont{Balseiro}}, \bibnamefont{and}
  \bibinfo{author}{\bibfnamefont{G.}~\bibnamefont{Usaj}},
  \bibinfo{journal}{{\it Unpublished}, arXiv:1409.2482v1}
  (\bibinfo{year}{2014}).

\bibitem[{\citenamefont{G\'omez-Le\'on and Platero}(2013)}]{Leon}
\bibinfo{author}{\bibfnamefont{A.}~\bibnamefont{G\'omez-Le\'on}}
  \bibnamefont{and} \bibinfo{author}{\bibfnamefont{G.}~\bibnamefont{Platero}},
  \bibinfo{journal}{Phys. Rev. Lett.} \textbf{\bibinfo{volume}{110}},
  \bibinfo{pages}{200403} (\bibinfo{year}{2013}).

\bibitem[{\citenamefont{Rudner et~al.}(2013)\citenamefont{Rudner, Lindner,
  Berg, and Levin}}]{Rudner}
\bibinfo{author}{\bibfnamefont{M.~S.} \bibnamefont{Rudner}},
  \bibinfo{author}{\bibfnamefont{N.~H.} \bibnamefont{Lindner}},
  \bibinfo{author}{\bibfnamefont{E.}~\bibnamefont{Berg}}, \bibnamefont{and}
  \bibinfo{author}{\bibfnamefont{M.}~\bibnamefont{Levin}},
  \bibinfo{journal}{Phys. Rev. X} \textbf{\bibinfo{volume}{3}},
  \bibinfo{pages}{031005} (\bibinfo{year}{2013}).

\bibitem[{\citenamefont{Kitagawa et~al.}(2010)\citenamefont{Kitagawa, Berg,
  Rudner, and Demler}}]{Kitagawa2}
\bibinfo{author}{\bibfnamefont{T.}~\bibnamefont{Kitagawa}},
  \bibinfo{author}{\bibfnamefont{E.}~\bibnamefont{Berg}},
  \bibinfo{author}{\bibfnamefont{M.}~\bibnamefont{Rudner}}, \bibnamefont{and}
  \bibinfo{author}{\bibfnamefont{E.}~\bibnamefont{Demler}},
  \bibinfo{journal}{Phys. Rev. B} \textbf{\bibinfo{volume}{82}},
  \bibinfo{pages}{235114} (\bibinfo{year}{2010}).

\bibitem[{\citenamefont{Kundu et~al.}(2014)\citenamefont{Kundu, Fertig, and
  Seradjeh}}]{Kundu2}
\bibinfo{author}{\bibfnamefont{A.}~\bibnamefont{Kundu}},
  \bibinfo{author}{\bibfnamefont{H.~A.} \bibnamefont{Fertig}},
  \bibnamefont{and} \bibinfo{author}{\bibfnamefont{B.}~\bibnamefont{Seradjeh}},
  \bibinfo{journal}{Phys. Rev. Lett.} \textbf{\bibinfo{volume}{113}},
  \bibinfo{pages}{236803} (\bibinfo{year}{2014}).

\bibitem[{\citenamefont{Tenenbaum~Katan and Podolsky}(2013)}]{Katan}
\bibinfo{author}{\bibfnamefont{Y.}~\bibnamefont{Tenenbaum~Katan}}
  \bibnamefont{and} \bibinfo{author}{\bibfnamefont{D.}~\bibnamefont{Podolsky}},
  \bibinfo{journal}{Phys. Rev. B} \textbf{\bibinfo{volume}{88}},
  \bibinfo{pages}{224106} (\bibinfo{year}{2013}).

\bibitem[{\citenamefont{Jiang et~al.}(2011)\citenamefont{Jiang, Kitagawa,
  Alicea, Akhmerov, Pekker, Refael, Cirac, Demler, Lukin, and Zoller}}]{Jiang}
\bibinfo{author}{\bibfnamefont{L.}~\bibnamefont{Jiang}},
  \bibinfo{author}{\bibfnamefont{T.}~\bibnamefont{Kitagawa}},
  \bibinfo{author}{\bibfnamefont{J.}~\bibnamefont{Alicea}},
  \bibinfo{author}{\bibfnamefont{A.~R.} \bibnamefont{Akhmerov}},
  \bibinfo{author}{\bibfnamefont{D.}~\bibnamefont{Pekker}},
  \bibinfo{author}{\bibfnamefont{G.}~\bibnamefont{Refael}},
  \bibinfo{author}{\bibfnamefont{J.~I.} \bibnamefont{Cirac}},
  \bibinfo{author}{\bibfnamefont{E.}~\bibnamefont{Demler}},
  \bibinfo{author}{\bibfnamefont{M.~D.} \bibnamefont{Lukin}}, \bibnamefont{and}
  \bibinfo{author}{\bibfnamefont{P.}~\bibnamefont{Zoller}},
  \bibinfo{journal}{Phys. Rev. Lett.} \textbf{\bibinfo{volume}{106}},
  \bibinfo{pages}{220402} (\bibinfo{year}{2011}).

\bibitem[{\citenamefont{Kundu and Seradjeh}(2013)}]{Kundu}
\bibinfo{author}{\bibfnamefont{A.}~\bibnamefont{Kundu}} \bibnamefont{and}
  \bibinfo{author}{\bibfnamefont{B.}~\bibnamefont{Seradjeh}},
  \bibinfo{journal}{Phys. Rev. Lett.} \textbf{\bibinfo{volume}{111}},
  \bibinfo{pages}{136402} (\bibinfo{year}{2013}).

\bibitem[{\citenamefont{Liu et~al.}(2013)\citenamefont{Liu, Levchenko, and
  Baranger}}]{Liu}
\bibinfo{author}{\bibfnamefont{D.~E.} \bibnamefont{Liu}},
  \bibinfo{author}{\bibfnamefont{A.}~\bibnamefont{Levchenko}},
  \bibnamefont{and} \bibinfo{author}{\bibfnamefont{H.~U.}
  \bibnamefont{Baranger}}, \bibinfo{journal}{Phys. Rev. Lett.}
  \textbf{\bibinfo{volume}{111}}, \bibinfo{pages}{047002}
  (\bibinfo{year}{2013}).

\bibitem[{\citenamefont{We et~al.}(2013)\citenamefont{We, Sun, Huang, Li, and
  Liu}}]{Wu}
\bibinfo{author}{\bibfnamefont{C.}~\bibnamefont{We}},
  \bibinfo{author}{\bibfnamefont{J.}~\bibnamefont{Sun}},
  \bibinfo{author}{\bibfnamefont{F.}~\bibnamefont{Huang}},
  \bibinfo{author}{\bibfnamefont{Y.}~\bibnamefont{Li}}, \bibnamefont{and}
  \bibinfo{author}{\bibfnamefont{W.}~\bibnamefont{Liu}}, \bibinfo{journal}{EPL}
  \textbf{\bibinfo{volume}{104}}, \bibinfo{pages}{27004}
  (\bibinfo{year}{2013}).

\bibitem[{\citenamefont{Wang et~al.}(2014)\citenamefont{Wang, Sun, and
  Xie}}]{Wang2}
\bibinfo{author}{\bibfnamefont{P.}~\bibnamefont{Wang}},
  \bibinfo{author}{\bibfnamefont{Q.-f.} \bibnamefont{Sun}}, \bibnamefont{and}
  \bibinfo{author}{\bibfnamefont{X.~C.} \bibnamefont{Xie}},
  \bibinfo{journal}{Phys. Rev. B} \textbf{\bibinfo{volume}{90}},
  \bibinfo{pages}{155407} (\bibinfo{year}{2014}).

\bibitem[{\citenamefont{Delplace et~al.}(2013)\citenamefont{Delplace,
  G\'omez-Le\'on, and Platero}}]{Delplace}
\bibinfo{author}{\bibfnamefont{P.}~\bibnamefont{Delplace}},
  \bibinfo{author}{\bibfnamefont{A.}~\bibnamefont{G\'omez-Le\'on}},
  \bibnamefont{and} \bibinfo{author}{\bibfnamefont{G.}~\bibnamefont{Platero}},
  \bibinfo{journal}{Phys. Rev. B} \textbf{\bibinfo{volume}{88}},
  \bibinfo{pages}{245422} (\bibinfo{year}{2013}).

\bibitem[{\citenamefont{Li et~al.}(2014)\citenamefont{Li, Kundu, Zhong, and
  Seradjeh}}]{Li4}
\bibinfo{author}{\bibfnamefont{Y.}~\bibnamefont{Li}},
  \bibinfo{author}{\bibfnamefont{A.}~\bibnamefont{Kundu}},
  \bibinfo{author}{\bibfnamefont{F.}~\bibnamefont{Zhong}}, \bibnamefont{and}
  \bibinfo{author}{\bibfnamefont{B.}~\bibnamefont{Seradjeh}},
  \bibinfo{journal}{Phys. Rev. B} \textbf{\bibinfo{volume}{90}},
  \bibinfo{pages}{121401} (\bibinfo{year}{2014}).

\bibitem[{\citenamefont{Dehghani et~al.}(2014)\citenamefont{Dehghani, Oka, and
  Mitra}}]{Dehghani1}
\bibinfo{author}{\bibfnamefont{H.}~\bibnamefont{Dehghani}},
  \bibinfo{author}{\bibfnamefont{T.}~\bibnamefont{Oka}}, \bibnamefont{and}
  \bibinfo{author}{\bibfnamefont{A.}~\bibnamefont{Mitra}},
  \bibinfo{journal}{Phys. Rev. B} \textbf{\bibinfo{volume}{90}},
  \bibinfo{pages}{195429} (\bibinfo{year}{2014}).

\bibitem[{\citenamefont{Dehghani et~al.}(2015)\citenamefont{Dehghani, Oka, and
  Mitra}}]{Dehghani2}
\bibinfo{author}{\bibfnamefont{H.}~\bibnamefont{Dehghani}},
  \bibinfo{author}{\bibfnamefont{T.}~\bibnamefont{Oka}}, \bibnamefont{and}
  \bibinfo{author}{\bibfnamefont{A.}~\bibnamefont{Mitra}},
  \bibinfo{journal}{Phys. Rev. B} \textbf{\bibinfo{volume}{91}},
  \bibinfo{pages}{155422} (\bibinfo{year}{2015}).

\bibitem[{\citenamefont{Titum et~al.}(2015)\citenamefont{Titum, Lindner,
  Rechtsman, and Refael}}]{paraj}
\bibinfo{author}{\bibfnamefont{P.}~\bibnamefont{Titum}},
  \bibinfo{author}{\bibfnamefont{N.~H.} \bibnamefont{Lindner}},
  \bibinfo{author}{\bibfnamefont{M.~C.} \bibnamefont{Rechtsman}},
  \bibnamefont{and} \bibinfo{author}{\bibfnamefont{G.}~\bibnamefont{Refael}},
  \bibinfo{journal}{Phys. Rev. Lett.} \textbf{\bibinfo{volume}{114}},
  \bibinfo{pages}{056801} (\bibinfo{year}{2015}).

\bibitem[{\citenamefont{Sambe}(1973)}]{Sambe}
\bibinfo{author}{\bibfnamefont{H.}~\bibnamefont{Sambe}},
  \bibinfo{journal}{Phys. Rev. A} \textbf{\bibinfo{volume}{7}},
  \bibinfo{pages}{2203} (\bibinfo{year}{1973}).

\bibitem[{\citenamefont{Rechtsman et~al.}(2013)\citenamefont{Rechtsman, Zeuner,
  Plotnik, Lumer, Podolsky, Dreisow, Nolte, Segev, and
  Szameit}}]{rechtsman2013photonic}
\bibinfo{author}{\bibfnamefont{M.~C.} \bibnamefont{Rechtsman}},
  \bibinfo{author}{\bibfnamefont{J.~M.} \bibnamefont{Zeuner}},
  \bibinfo{author}{\bibfnamefont{Y.}~\bibnamefont{Plotnik}},
  \bibinfo{author}{\bibfnamefont{Y.}~\bibnamefont{Lumer}},
  \bibinfo{author}{\bibfnamefont{D.}~\bibnamefont{Podolsky}},
  \bibinfo{author}{\bibfnamefont{F.}~\bibnamefont{Dreisow}},
  \bibinfo{author}{\bibfnamefont{S.}~\bibnamefont{Nolte}},
  \bibinfo{author}{\bibfnamefont{M.}~\bibnamefont{Segev}}, \bibnamefont{and}
  \bibinfo{author}{\bibfnamefont{A.}~\bibnamefont{Szameit}},
  \bibinfo{journal}{Nature} \textbf{\bibinfo{volume}{496}},
  \bibinfo{pages}{196} (\bibinfo{year}{2013}).

\bibitem[{\citenamefont{Sentef et~al.}(2015)\citenamefont{Sentef, Claassen,
  Kemper, Moritz, Oka, Freericks, and Devereaux}}]{sentef2015theory}
\bibinfo{author}{\bibfnamefont{M.}~\bibnamefont{Sentef}},
  \bibinfo{author}{\bibfnamefont{M.}~\bibnamefont{Claassen}},
  \bibinfo{author}{\bibfnamefont{A.}~\bibnamefont{Kemper}},
  \bibinfo{author}{\bibfnamefont{B.}~\bibnamefont{Moritz}},
  \bibinfo{author}{\bibfnamefont{T.}~\bibnamefont{Oka}},
  \bibinfo{author}{\bibfnamefont{J.}~\bibnamefont{Freericks}},
  \bibnamefont{and}
  \bibinfo{author}{\bibfnamefont{T.}~\bibnamefont{Devereaux}},
  \bibinfo{journal}{Nature communications} \textbf{\bibinfo{volume}{6}}
  (\bibinfo{year}{2015}).

\bibitem[{\citenamefont{Wilson et~al.}(2016)\citenamefont{Wilson, Song, and
  Refael}}]{Wilson}
\bibinfo{author}{\bibfnamefont{J.~H.} \bibnamefont{Wilson}},
  \bibinfo{author}{\bibfnamefont{J.~C.~W.} \bibnamefont{Song}},
  \bibnamefont{and} \bibinfo{author}{\bibfnamefont{G.}~\bibnamefont{Refael}},
  \bibinfo{journal}{{\it unpublished} arXiv:1603.01621}
  (\bibinfo{year}{2016}).

\bibitem[{\citenamefont{D'Alessio and Rigol}(2014{\natexlab{a}})}]{DAlessio}
\bibinfo{author}{\bibfnamefont{L.}~\bibnamefont{D'Alessio}} \bibnamefont{and}
  \bibinfo{author}{\bibfnamefont{M.}~\bibnamefont{Rigol}},
  \bibinfo{journal}{Phys. Rev. X} \textbf{\bibinfo{volume}{4}},
  \bibinfo{pages}{041048} (\bibinfo{year}{2014}{\natexlab{a}}).

\bibitem[{\citenamefont{D'Alessio and Rigol}(2014{\natexlab{b}})}]{DAlessio2}
\bibinfo{author}{\bibfnamefont{L.}~\bibnamefont{D'Alessio}} \bibnamefont{and}
  \bibinfo{author}{\bibfnamefont{M.}~\bibnamefont{Rigol}},
  \bibinfo{journal}{{\it Unpublished}, arXiv:1409.6319}
  (\bibinfo{year}{2014}{\natexlab{b}}).

\bibitem[{\citenamefont{Seetharam et~al.}(2015)\citenamefont{Seetharam, Bardyn,
  Lindner, Rudner, and Refael}}]{karthik}
\bibinfo{author}{\bibfnamefont{K.~I.} \bibnamefont{Seetharam}},
  \bibinfo{author}{\bibfnamefont{C.-E.} \bibnamefont{Bardyn}},
  \bibinfo{author}{\bibfnamefont{N.~H.} \bibnamefont{Lindner}},
  \bibinfo{author}{\bibfnamefont{M.~S.} \bibnamefont{Rudner}},
  \bibnamefont{and} \bibinfo{author}{\bibfnamefont{G.}~\bibnamefont{Refael}},
  \bibinfo{journal}{{\it Unpublished}, arXiv:1502.02664}
  (\bibinfo{year}{2015}).

\bibitem[{\citenamefont{Lazarides et~al.}(2014)\citenamefont{Lazarides, Das,
  and Moessner}}]{PhysRevE.90.012110}
\bibinfo{author}{\bibfnamefont{A.}~\bibnamefont{Lazarides}},
  \bibinfo{author}{\bibfnamefont{A.}~\bibnamefont{Das}}, \bibnamefont{and}
  \bibinfo{author}{\bibfnamefont{R.}~\bibnamefont{Moessner}},
  \bibinfo{journal}{Phys. Rev. E} \textbf{\bibinfo{volume}{90}},
  \bibinfo{pages}{012110} (\bibinfo{year}{2014}),
  \urlprefix\url{http://link.aps.org/doi/10.1103/PhysRevE.90.012110}.

\bibitem[{\citenamefont{Ponte et~al.}(2015)\citenamefont{Ponte, Chandran,
  Papić, and Abanin}}]{Ponte}
\bibinfo{author}{\bibfnamefont{P.}~\bibnamefont{Ponte}},
  \bibinfo{author}{\bibfnamefont{A.}~\bibnamefont{Chandran}},
  \bibinfo{author}{\bibfnamefont{Z.}~\bibnamefont{Papić}}, \bibnamefont{and}
  \bibinfo{author}{\bibfnamefont{D.~A.} \bibnamefont{Abanin}},
  \bibinfo{journal}{Annals of Physics} \textbf{\bibinfo{volume}{353}},
  \bibinfo{pages}{196 } (\bibinfo{year}{2015}), ISSN \bibinfo{issn}{0003-4916},
  \urlprefix\url{http://www.sciencedirect.com/science/article/pii/S0003491614003212}.

\bibitem[{\citenamefont{Tien and Gordon}(1963)}]{TienGordon}
\bibinfo{author}{\bibfnamefont{P.~K.} \bibnamefont{Tien}} \bibnamefont{and}
  \bibinfo{author}{\bibfnamefont{J.~P.} \bibnamefont{Gordon}},
  \bibinfo{journal}{Phys. Rev.} \textbf{\bibinfo{volume}{129}},
  \bibinfo{pages}{647} (\bibinfo{year}{1963}).

\bibitem[{\citenamefont{Freericks et~al.}(2009)\citenamefont{Freericks,
  Krishnamurthy, and Pruschke}}]{freericks2009theoretical}
\bibinfo{author}{\bibfnamefont{J.}~\bibnamefont{Freericks}},
  \bibinfo{author}{\bibfnamefont{H.}~\bibnamefont{Krishnamurthy}},
  \bibnamefont{and} \bibinfo{author}{\bibfnamefont{T.}~\bibnamefont{Pruschke}},
  \bibinfo{journal}{Physical review letters} \textbf{\bibinfo{volume}{102}},
  \bibinfo{pages}{136401} (\bibinfo{year}{2009}).

\bibitem[{\citenamefont{Freericks et~al.}(2014)\citenamefont{Freericks,
  Krishnamurthy, Sentef, and Devereaux}}]{freericks2014gauge}
\bibinfo{author}{\bibfnamefont{J.}~\bibnamefont{Freericks}},
  \bibinfo{author}{\bibfnamefont{H.}~\bibnamefont{Krishnamurthy}},
  \bibinfo{author}{\bibfnamefont{M.}~\bibnamefont{Sentef}}, \bibnamefont{and}
  \bibinfo{author}{\bibfnamefont{T.}~\bibnamefont{Devereaux}},
  \bibinfo{journal}{arXiv preprint arXiv:1403.7585}  (\bibinfo{year}{2014}).

\bibitem[{\citenamefont{Fregoso et~al.}(2013)\citenamefont{Fregoso, Wang,
  Gedik, and Galitski}}]{fregoso}
\bibinfo{author}{\bibfnamefont{B.~M.} \bibnamefont{Fregoso}},
  \bibinfo{author}{\bibfnamefont{Y.}~\bibnamefont{Wang}},
  \bibinfo{author}{\bibfnamefont{N.}~\bibnamefont{Gedik}}, \bibnamefont{and}
  \bibinfo{author}{\bibfnamefont{V.}~\bibnamefont{Galitski}},
  \bibinfo{journal}{Physical Review B} \textbf{\bibinfo{volume}{88}},
  \bibinfo{pages}{155129} (\bibinfo{year}{2013}).

\bibitem[{\citenamefont{Fu}(2009)}]{FuWarp}
\bibinfo{author}{\bibfnamefont{L.}~\bibnamefont{Fu}}, \bibinfo{journal}{Phys.
  Rev. Lett.} \textbf{\bibinfo{volume}{103}}, \bibinfo{pages}{266801}
  (\bibinfo{year}{2009}),
  \urlprefix\url{http://link.aps.org/doi/10.1103/PhysRevLett.103.266801}.

\bibitem[{\citenamefont{Marchand and Franz}(2012)}]{Marchand}
\bibinfo{author}{\bibfnamefont{D.~J.~J.} \bibnamefont{Marchand}}
  \bibnamefont{and} \bibinfo{author}{\bibfnamefont{M.}~\bibnamefont{Franz}},
  \bibinfo{journal}{Phys. Rev. B} \textbf{\bibinfo{volume}{86}},
  \bibinfo{pages}{155146} (\bibinfo{year}{2012}),
  \urlprefix\url{http://link.aps.org/doi/10.1103/PhysRevB.86.155146}.

\end{thebibliography}

\appendix 
\section{Integrals Involving the Pump Envelope}\label{ap:PumpEnvelope}
As discussed above, we choose to describe this electric field in a gauge where the scalar potential is zero. Thus we have
\beqa
&&\Av_{pump}(t) = - \int_{-\infty}^t dt' {\bf E}_{pump}(t') \\ \nonumber &=& -E_0  \int_{-\infty}^t dt' e^{-\frac{t'^2}{2T_{pump}^2}} {\bf E}_{\Omega}(t')
\eeqa
where we have chosen in initial condition such that $\Av_{pump}(t)\to0$ for $t\to-\infty$. 

Let us define the frequency scale associated with the pump pulse $\Omega_{pump}=2\pi/T_{pump}$. We work in the limit
\beq
\Omega_{pump}\ll \Omega
\eeq
such that there are many oscillations within the pump field envelope. 
We now define $ {\bf E}_{\Omega}(t)=-\frac{ \tilde{\bf E}_{\Omega}'(t)}{\Omega}$ and integrate by parts to obtain
\beqa
\Av_{pump}(t) &=& \frac{E_0}{\Omega} e^{-\frac{t'^2}{2T_{pump}^2}} \tilde{\bf E}_{\Omega}(t')\big{|}_{-\infty}^{t} \\ \nonumber &+&\frac{E_0}{ \Omega T_{pump}^2}  \int_{-\infty}^t dt' t'e^{-\frac{t'^2}{2T_{pump}^2}} \tilde{\bf E}_{\Omega}(t')\\ \nonumber &=& \frac{E_0}{\Omega} e^{-\frac{t^2}{2T_{pump}^2}} \tilde{\bf E}_{\Omega}(t) +\mathcal{O}\left(\frac{\Omega_{pump}}{\Omega}\right)
\eeqa

The above process could in principle be iterated to produce a perturbative expansion in $\frac{\Omega_{pump}}{\Omega}$, although we stop here for practicality. We could alternatively write $\frac{\Omega_{pump}}{\Omega}=\frac{T}{T_{pump}}$, which tells us this expression is valid in the limit $T\ll T_{probe}$; i.e. the amplitude changes on a much longer time scale than the period of oscillation. We neglect all but the leading order terms. Continuing the procedure above shows that the next to leading order term is of order $\left(\frac{\Omega_{pump}}{\Omega}\right)^2$. Thus we work in a regime where
\beq
\Av_{pump}(t)\simeq \frac{E_0}{\Omega} e^{-\frac{t^2}{2T_{pump}^2}} \tilde{\bf E}_{\Omega}(t)
\eeq

\section{Green's Function}\label{ap:GF}
We now consider the quantity
\beq
G^{<}_{\kv, \alpha\beta}(t,t')=i\langle c^\dagger_{\kv\beta}(t') c_{\kv\alpha}(t)\rangle
\eeq
In order to define a useful quantity we consider the equation of motion for the electronic operators:
\beq
\dot{c}_{\kv\alpha}(t)=i[\mathcal{H}(t), c_{\kv\alpha}(t)]
\eeq
where the over-dot denotes differentiation with respect to time and $\mathcal{H}(t)= \sum_{\kv,\alpha,\beta} c_{\kv\alpha}^\dagger H_{\kv,\alpha\beta}(t) c_{\kv\beta}$. Using the Hamiltonian defined above and calculating the commutator gives
\beq
\dot{c}_{\kv\alpha}(t)=-i H_{\kv,\alpha\beta}(t) c_{\kv\beta}(t)
\eeq
where summation over repeated indices is implied. The above equation is linear in electron operators. We thus try a solution of the form $c_{\kv\alpha}(t)= U_{\kv\alpha\alpha'}(t,t_r)c_{\kv\alpha'}(t_r)$ where the $U_{\kv\alpha\alpha'}(t,t_r)$ are complex numbers. Plugging this in gives
\beq
\dot{U}_{\kv\alpha\alpha'}(t,t_r)c_{\kv\alpha'}(t_r)=-i H_{\kv,\alpha\beta}(t)  U_{\kv\beta\alpha'}(t,t_r)c_{\kv\alpha'}(t_r)
\eeq
Which implies
\beq
i\dot{U}_{\kv\alpha\alpha'}(t,t_r)=H_{\kv,\alpha\beta}(t) U_{\kv\beta\alpha'}(t,t_r)
\eeq
promoting $U$ and $H$ to matrices gives
\beq
i\partial_t{U}_{\kv}(t,t_r)=H_{\kv}(t)  U_{\kv}(t,t_r)
\eeq
The formal solution to the above equation is
\beq
{U}_{\kv}(t,t_r)=T\left(e^{-i\int_{t_r}^td\tau H_{\kv}(t) }\right)
\eeq
and it obeys ${U}_{\kv}(t,t'){U}_{\kv}(t',t_r)={U}_{\kv}(t,t_r)$ and $({U}_{\kv}(t,t_r))^\dagger = {U}_{\kv}(t_r,t)$, where $T$ is the time ordering operator.
Using this solution we can write
\beqa
&& G^{<}_{\kv, \alpha\beta}(t,t')=i U_{\kv\alpha\alpha'}(t,t_r)U^*_{\kv\beta\beta'}(t',t_r)  \langle c^\dagger_{\kv\beta'}(t_r) c_{\kv\alpha'}(t_r)\rangle\nonumber \\ 
&=& U_{\kv\alpha\alpha'}(t,t_r)G^{<}_{\kv, \alpha'\beta'}(t_r,t_r)U^*_{\kv\beta\beta'}(t',t_r)
\eeqa
Writing the above in matrix form gives
\beqa
G^{<}_{\kv}(t,t')&=& U_{\kv}(t,t_r)G^{<}_{\kv}(t_r,t_r)U^\dagger_{\kv}(t',t_r)
\eeqa
or
\beqa
G^{<}_{\kv}(t,t')&=& U_{\kv}(t,t_r)G^{<}_{\kv}(t_r,t_r)U_{\kv}(t_r, t')
\eeqa
Which is conceptually appealing. We begin at $t'$, propagate back to $t_r$ where we know the Green's function and then propagate forward to $t$. We now assume that the system begins at time $t_r$ in equilibrium in a system obeying the unperturbed Hamiltonian. Thus we write $G^{<}_{\kv}(t_r,t_r)=i\sum_{\alpha} |\phi_{\kv\alpha}\rangle\langle \phi_{\kv\alpha}| f(E_{\kv\alpha})$. Noting that $|\psi_{\kv\alpha}(t)\rangle = U_{\kv}(t,t_r) |\phi_{\kv\alpha}\rangle$ then immediately leads to the expression for the Green's function used in the main text, Eq.~\ref{eq:GLesser} .

\section{Gauge Choice}\label{ap:Gauge}
In considering the effects of electromagnetic fields we must ensure that our theory is gauge invariant. A general gauge transformation is applied as follows 
\beqa
&&\Av(\rv,t)\to  \Av(\rv,t)+\nabla \chi(\rv,t)\\ \nonumber
&&\Phi(\rv,t)\to  \Phi(\rv,t)-\partial_t\chi(\rv,t)\\ \nonumber
&&c_{j\sigma} \to e^{ie\chi(\rv_j,t)/\hbar}c_{j\sigma}  \
\eeqa
Within the so called ``Hamiltonian gauge" used above we take $\Phi=0$ and ${\bf E}(t)=-\partial_t \Av(t)$. Therefore, to remain within this choice of Gauge and not change our problem in a non-trivial way by for example, introducing a spatial dependence, we must be free to introduce a Gauge change $\chi = \rv \cdot {\bf f}$ where ${\bf f}$ is an arbitrary, {\em constant} vector. This amounts to the Gauge change $\Av(t)\to  \Av(t)+{\bf f}$, $\Phi(\rv,t)\to  \Phi(\rv,t)$ and $c_{j\sigma} \to e^{ie\rv_j \cdot {\bf f} /\hbar}c_{j\sigma}$. Fourier transforming the electron annihilation operator leads to the result that the Gauge change makes the modification $c_{\kv\sigma}\to c_{\kv-e{\bf f}, \sigma}$. Note also that $H_{\kv}(t)\to H_{\kv-e{\bf f}}(t)$ under this transformation. Thus our time evolution operators change as $U_{\kv}(t,t')\to U_{\kv-e{\bf f}}(t,t')$ and therefore all of the Greens functions defined above transform as $G_{\kv}(t,t')\to G_{\kv-e{\bf f}}(t,t')$ and are thus not Gauge invariant.

It is useful to note conceptually where this Gauge freedom comes from. We require ${\bf E}(t)=-\partial_t \Av(t)$ which in turn gives
\beq
\Av(t)= -\int_{\mathcal{O}}^ t dt' {\bf E}(t')+\Av(\mathcal{O})
\eeq
where the initial condition $\Av(\mathcal{O})$ is {\em unfixed} by the electric field. Thus the freedom we have lies in our choice of the reference $\Av(\mathcal{O})$. Recall the turn on procedure we have in mind: the pump field is off for $t<t_r$ and is switched on after-words. Therefore, for this choice of set-up it is convenient to express $\Av$ as follows
\beqa
\Av(t)&=& -\Theta(t-t_0)\int_{t_r}^ t dt' {\bf E}(t')+\Av(t_0)\\ \nonumber &=&\Av_{physical}(t)+\Av(t_r)
\eeqa
but $\Av(t_r)\equiv \Av_0$ is still entirely arbitrary. Thus we would like a theory completely independent of $\Av(t_r)$. This is equivalent to the statement above that the Gauge invariant quantities should be independent of ${\bf f}$, as all ${\bf f}$ constitutes is a shift in the value of $\Av(t_r)$.

We now note that Green's functions $\tilde{G}_{\kv}(t,t')\equiv G_{\kv+e\Av_0}(t,t')$ are unchanged by the Gauge transformation $\Av(t)\to \Av(t)+{\bf f}$ as the shift $\kv\to \kv-e{\bf f}$ in the Gauge dependent wave function cancels out the shift $\Av_0\to \Av_0+{\bf f}$. An equivalent finding is that
\beq
{U}_{\kv+e\Av_0}(t,t_r)=T\left(e^{-i\int_{t_r}^td\tau H_{\kv+e\Av_0}(t)}\right)
\eeq
but
\beqa
&& H_{\kv+e\Av_0}(t)= h_{\kv -e\Av(t)+e\Av_0}\\ \nonumber &&= h_{\kv -e\Av_{physical}(t)-e\Av_0+e\Av_0}= h_{\kv -e\Av_{physical}(t)}
\eeqa
where $h_{\kv}$ is the Dirac Hamiltonian. Therefore $H_{\kv+e\Av_0}(t)$ is independent of our arbitrary choice of $\Av_0$, meaning that ${U}_{\kv+e\Av_0}(t,t_r)$ and thus $ G_{\kv+e\Av_0}(t,t')$ are gauge invariant as well. 

Of course the most natural choice is to set $\Av_0=0$ so that the (equilibrium) Hamiltonian before the switch on time $t_r$ is simply $h_\kv$, as one would like it to be. For {\em this choice} of initial condition the Gauge invariant Greens function and the traditional Greens function are identical.

\section{Linearly Polarized Light}\label{ap:Linearly}
Here we go over the details leading to Eq.~(\ref{Linearfinal}) in the main text. Assuming $T_{pump}\ll 2\pi/\omega$, $$\int_{t'}^t dt'' A_x(t'') =  \frac{E_0}{\Omega^2} \left(e^{-t^2/2T_{pump}^2} \sin{\Omega t}-e^{-t'^2/2T_{pump}^2} \sin{\Omega t'}\right)$$ and using the identity  $e^{ix\sin{\Omega t}}=\sum_{m} J_m(x)e^{im\Omega t}$ in the wave functions for the linarly polarized light the photocurrent with $k_y=0$ reads
\begin{widetext}
\beqa
I(k_x,0,\omega, t_{\mathcal{O}})=\sum_{\alpha} f(\epsilon_{k_x,0,\alpha})  \left| \sum_{m}   \int_{-\infty}^\infty dt_1  J_{m}\left({\alpha  \mathcal{A}(t_1)}{}\right)   e^{-\frac{(t_1-t_{\mathcal{O}})^2}{2T_{probe}^2}} e^{-i(\omega-\alpha v_Fk_x+\frac{\mu}{\hbar}-m\Omega) t_1}  \right|^2
\eeqa
\end{widetext}
In the above,  $ J_{m}\left({\alpha  \mathcal{A}(t_1)}\right)$ describes a splitting of the eigenstates into sidebands, labeled by $m$, where the amplitudes of these sidebands depend on time. Meanwhile $s(t_1-t_{\mathcal{0}})= e^{-\frac{(t_1-t_{\mathcal{O}})^2}{2T_{probe}^2}} $ is the profile of the probe pulse and $e^{-i(\omega+\alpha v_Fk_x+\frac{\mu}{\hbar}-m\Omega) t_1} $ describes having energies not just at $\pm v_Fk_x-\mu$, but also at values $m\hbar\Omega$ above and below these values.
The integral above is of course intractable to perform exactly. We can make progress using a series of appropriate approximations. The first is that the probe pulse is much shorter than the pump pulse. Therefore $\mathcal{A}(t)$ changes very slowly over the duration of $ e^{-\frac{(t_1-t_{\mathcal{O}})^2}{2T_{probe}^2}} $. We can therefore simply replace the $J_{m}\left({\alpha  \mathcal{A}(t_1)}\right)$ term with its value at the peak $t_1=t_{\mathcal{O}}$. However we observe that a better approach is to replace $\mathcal{A}(t)= \frac{e E_0v_F}{\hbar \Omega^2}e^{-t^2/2T_{pump}^2}$ with a weighted average over the probe pulse. Thus we define
\beq
A_{eff}(t_{\mathcal{O}}) = \frac{\int dt e^{-\frac{(t-t_{\mathcal{O}})^2}{2T_{probe}^2}}  \mathcal{A}(t)}{\int dt e^{-\frac{(t-t_{\mathcal{O}})^2}{2T_{probe}^2}} }
\eeq
Once we have made this replacement the remaining integral can be done analytically and gives
\begin{widetext}
\beqa
I(k_x, 0, \omega, t_{\mathcal{O}})&=&2\pi T_{probe}^2 \sum_{\alpha, m.m'}   f(\epsilon_{k_x,0,\alpha})J_{m}\left({\alpha  A_{eff}(t_{\mathcal{O}})}\right)J_{m'}\left({\alpha  A_{eff}(t_{\mathcal{O}})}{}\right)    \\ \nonumber
&\times& \exp\left[-(\omega-\alpha v_Fk_x+\frac{\mu}{\hbar}-m\Omega)^2T_{probe}^2/2\right]    \exp\left[-(\omega-\alpha v_Fk_x+\frac{\mu}{\hbar}-m'\Omega)^2T_{probe}^2/2\right]
\eeqa
\end{widetext}
The width of the peaks described by the Gaussians above are set by the frequency scale $1/T_{probe}$. The two Gaussians describe peaks centred at $\alpha v_Fk_x-\frac{\mu}{\hbar}+m\Omega$ and $\alpha v_Fk_x-\frac{\mu}{\hbar}+m'\Omega$. Thus the separation between the two peaks is $(m-m')\Omega$. If the decay scale $1/T_{probe}$ is much smaller than the smallest separation $\Omega$, i.e. $1/T_{probe}\ll \Omega$ then the peaks do not overlap at the same frequency and so the major contribution to the double sum comes from the $m'=m$ terms. Working in this approximation gives
\beqa
&&I(k_x, 0, \omega, t_{\mathcal{O}})=2\pi T_{probe}^2 \sum_{\alpha, m}  f(\epsilon_{k_x,0,\alpha})J_{m}^2\left({ A_{eff}(t_{\mathcal{O}})}\right)\nonumber  \\ &\times&\exp\left[-(\omega-\alpha v_Fk_x+\frac{\mu}{\hbar}-m\Omega)^2T_{probe}^2\right]
\eeqa

\section{Circularly Polarized Light}\label{ap:Circularly} 
\subsection{Analytic Work}
We begin with the approximate wave function found in the main text, reproduced here for convenience
\beqa
&&|\psi_{\Gamma \alpha}(t)\rangle = e^{i\mu(t-t_r)/\hbar} e^{i\Omega t \sigma_z/2}\\ \nonumber &\times& e^{-iH_{eff}(t-t_r)/\hbar} e^{-i\Omega t_r \sigma_z/2} |\phi_{\Gamma \alpha}\rangle
\eeqa
.  We note that in order to find the trace of the  Green's function we will require $\langle \phi_{\Gamma,\beta}|\psi_{\Gamma \alpha}(t)\rangle $ (and its complex conjugate). In the distant past the field is turned off and so the Hamiltonian is $h_\Gamma(t\to-\infty)=-\mu\sigma_0$. Thus we are free to choose any ``initial" set of states, provided they are orthonormal. For convenience we choose $\{|\phi_{\Gamma \alpha}\rangle\}$  to be $(1,0)^T$ and $(0,1)^T$, which we label  $|\phi_{\alpha}\rangle$ with $\alpha=\pm1$. From this point forward we will drop the $\Gamma$ subscript in the interest of brevity. This leads to the following
 \beqa
&&\langle \phi_{\Gamma,\beta}|\psi_{\Gamma \alpha}(t)\rangle = e^{i(\mu+\beta\hbar\Omega/2) t/\hbar}  e^{-i(\mu+\alpha\hbar\Omega/2)t_r/\hbar}\nonumber \\  &\times& \left \langle \phi_{\beta}\left |e^{-iH_{eff}(t-t_r)/\hbar} \right |\phi_{\alpha}\right\rangle
\eeqa
In order to calculate the matrix elements of $e^{-iH_{eff}(t-t_r)/\hbar} $ we write the argument of the exponential as follows
\beqa
\left( - \hbar v_FA_{eff}(t_{\mathcal{O}}) \sigma_y+\frac{\hbar\Omega}{2} \sigma_z  \right)= \frac{E_{eff}(t_{\mathcal{O}})}{\hbar} \hat{a} \cdot \vec{\sigma}
\eeqa
 where $\hat{a}=(- \frac{\hbar v_FA_{eff}(t_{\mathcal{O}})}{ E_{eff}(t_{\mathcal{O}})}\hat{y}+\frac{\frac{\hbar\Omega}{2} }{ E_{eff}(t_{\mathcal{O}})}\hat{z})$ is a unit vector and we remind the reader that $E_{eff}= \sqrt{(\hbar v_F A_{eff}(t_\mathcal{O}))^2+(\hbar\Omega/2)^2} $. Making use of the identity $e^{-i x \hat{a} \cdot \vec{\sigma}}= \cos\left(x\right) \sigma_0-i\sin\left( x\right)\hat{a} \cdot \vec{\sigma}$ one can show that
 \beqa
 && \left \langle \phi_{\beta}\left | e^{-iH_{eff}(t-t_r)/\hbar} \right |\phi_{\alpha}\right\rangle \nonumber \\  &=&\sum_{s=\pm1} e^{i \frac{s E_{eff}(t_{\mathcal{O}})}{\hbar}(t-t_r)}A^{s}_{\alpha,\beta}
 \eeqa
 where $A^{s}_{\alpha,\beta}= \left( \delta_{\alpha,\beta}-s \hat{a} \cdot    \left \langle \psi_{\beta}\left |\vec{\sigma} \right |\psi_{\alpha}\right\rangle\right)/2$ which leads to
 \begin{widetext}
  \beqa
\langle \phi_{\Gamma,\beta}|\psi_{\Gamma \alpha}(t)\rangle =\sum_{s=\pm1} e^{i(\mu+\beta\hbar\Omega/2+sE_{eff}(t_{\mathcal{O}})) t/\hbar}  e^{-i(\mu+\alpha\hbar\Omega/2+sE_{eff}(t_{\mathcal{O}}))t_r/\hbar}A^{s}_{\alpha,\beta}
\eeqa
\end{widetext}
Using the above, an equivalent result for $\langle \psi_{\Gamma\alpha}(t')|\phi_{\Gamma\beta}\rangle$, and performing the $t_1$ and $t_2$ integrals gives
\beqa
&&I(0,0,\omega, t_{\mathcal{O}})=\\ \nonumber && 2\pi T_{probe}^2 \sum_{\alpha\beta, s,s'} f(\epsilon_{\Gamma\alpha})e^{-i(s-s')E_{eff}(t_{\mathcal{O}})t_r/\hbar} A^{s}_{\alpha,\beta}(A^{s'}_{\alpha,\beta})^* \\\nonumber &\times& \exp\left[-(\omega+\mu/\hbar-\beta\Omega/2-sE_{eff}(t_{\mathcal{O}})/\hbar)^2T_{probe}^2/2\right]\\ \nonumber &\times&   \exp\left[-(\omega-\mu/\hbar-\beta\Omega/2-s'E_{eff}(t_{\mathcal{O}})/\hbar)^2T_{probe}^2/2\right]
\eeqa
Recalling that we are working under the assumption $E_{eff}(t_{\mathcal{O}})=\sqrt{(\hbar v_FA_{eff}(t_{\mathcal{O}}))^2+(\frac{\hbar\Omega}{2})^2}\ge \hbar\Omega/2 \gg1/T_{probe}$, the distance separating the peaks in the Gaussians above (which is $2E_{eff}(t_{\mathcal{O}})/\hbar$) is much large than the width of the peaks. We therefore discard terms where $s\ne s'$ which gives us
 \beqa
&&I(0,0,\omega, t_{\mathcal{O}})=2\pi T_{probe}^2 \sum_{\alpha\beta, s} f(\epsilon_{\Gamma\alpha}) |A^{s}_{\alpha,\beta}|^2 \\\nonumber &\times& \exp\left[-(\omega+\mu/\hbar-\beta\Omega/2-sE_{eff}(t_{\mathcal{O}})/\hbar)^2T_{probe}^2\right]
\eeqa
 We note that the eigenvalues in the distant past are $\epsilon_{\Gamma\alpha}=-\mu$, where are independent of $\alpha$ (as we're at the Dirac point). At this point the only $\alpha$ dependence left in the summand comes from the matrix elements $ |A^{s}_{\alpha,\beta}|^2$. Some algebra shows $\sum_{\alpha} |A^{s}_{\alpha,\beta}|^2  =\frac{1-s\beta \hat{a}_z}{2}$ leading to our final result
  \beqa
&&I(0,0,\omega, t_{\mathcal{O}})=2\pi T_{probe}^2  f(-\mu) \sum_{\beta, s} \left(\frac{1-s\beta \hat{a}_z}{2}\right)\\ \nonumber &\times&\exp\left[-(\omega+\mu/\hbar-\beta\Omega/2-sE_{eff}(t_{\mathcal{O}})/\hbar)^2T_{probe}^2\right]
\eeqa

\subsection{Beyond the Dirac Cone Model}
\begin{figure}[]
  \setlength{\unitlength}{1mm}

   \includegraphics[scale=.45]{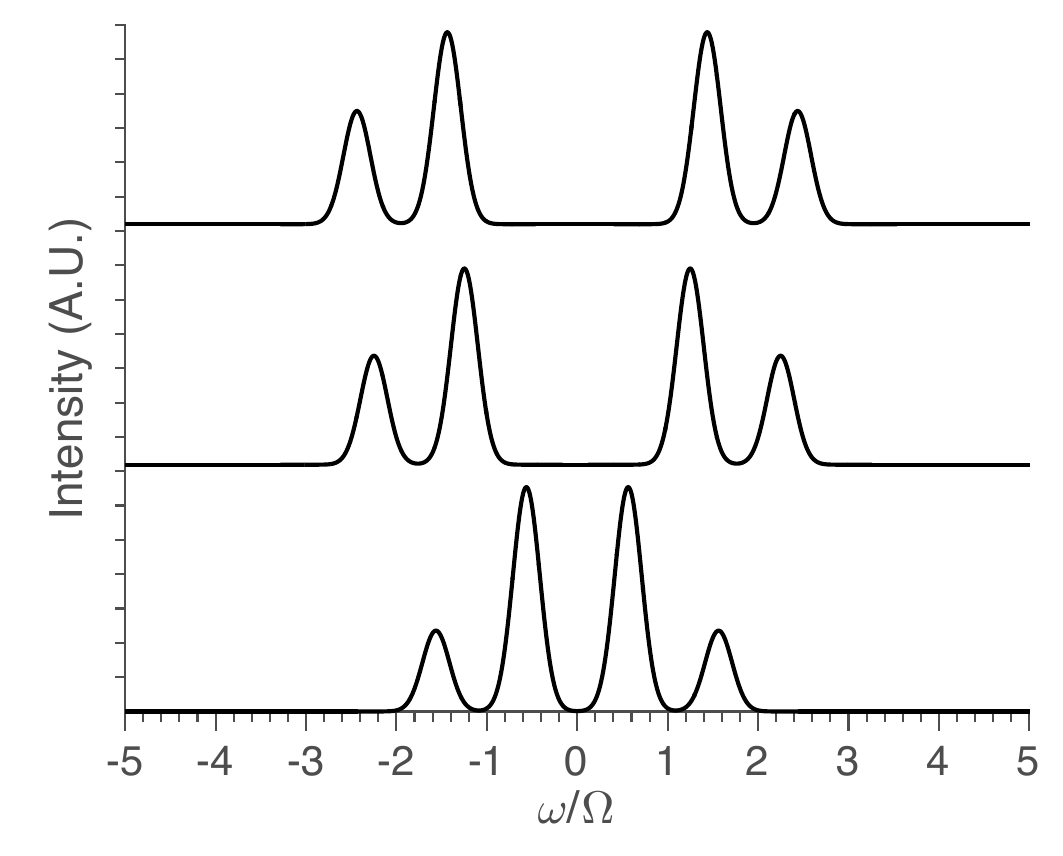}
\caption{{\small Plot of $I(0,0,\omega,t_\mathcal{O})$ at various delay times using the model Eq. (\ref{warped}). In these plots the bottom plot is for $t_\mathcal{O}=-500$fs, the middle for $t_\mathcal{O}=-100$fs and the top is for $t_\mathcal{O}=0$fs. Note the lack of any additional side-bands
     }
     }\label{fig:hexagonal_warping}
\end{figure}

One of the main results of this work is that the Dirac cone, under the application of circularly polarized light, admits only two side-bands at the Dirac point. Here we go beyond the Dirac cone description in order to argue that this behavior is universal and not a peculiarity of the linearized Dirac cone model. Towards this end we have added so-called hexgonal warping and velocity renormalization terms to our model\cite{FuWarp}, our equilibrium system then reads
\beq\label{warped}
h_{\kv} =  (\hbar v_F+ \Lambda k^2) (\kv\times\vec{\sigma})+\frac{\lambda}{2}(k_+^3+k_-^3)\sigma_z -\mu\sigma_0
\eeq
where $\Lambda$ and $\lambda$ characterize the hexagonal warping and velocity renormalization terms respectively. In the above we have defined $k_{\pm}=k_x\pm ik_y$. We use as estimates for these parameter values estimated from experiment, $\lambda=50eV\AA^3$ and $\Lambda=100eV\AA^3$. Our numerical results are shown in Fig \ref{fig:hexagonal_warping}. We see that the inclusion of these terms does not change the conclusion that only two side-bands exist. We have run simulations at $\lambda$ and $\Lambda$ values two orders of magnitude larger than the physical values above and still find the same behaviour.

\begin{widetext}
 \begin{figure*}[]
  	\includegraphics[width = 0.8\columnwidth]{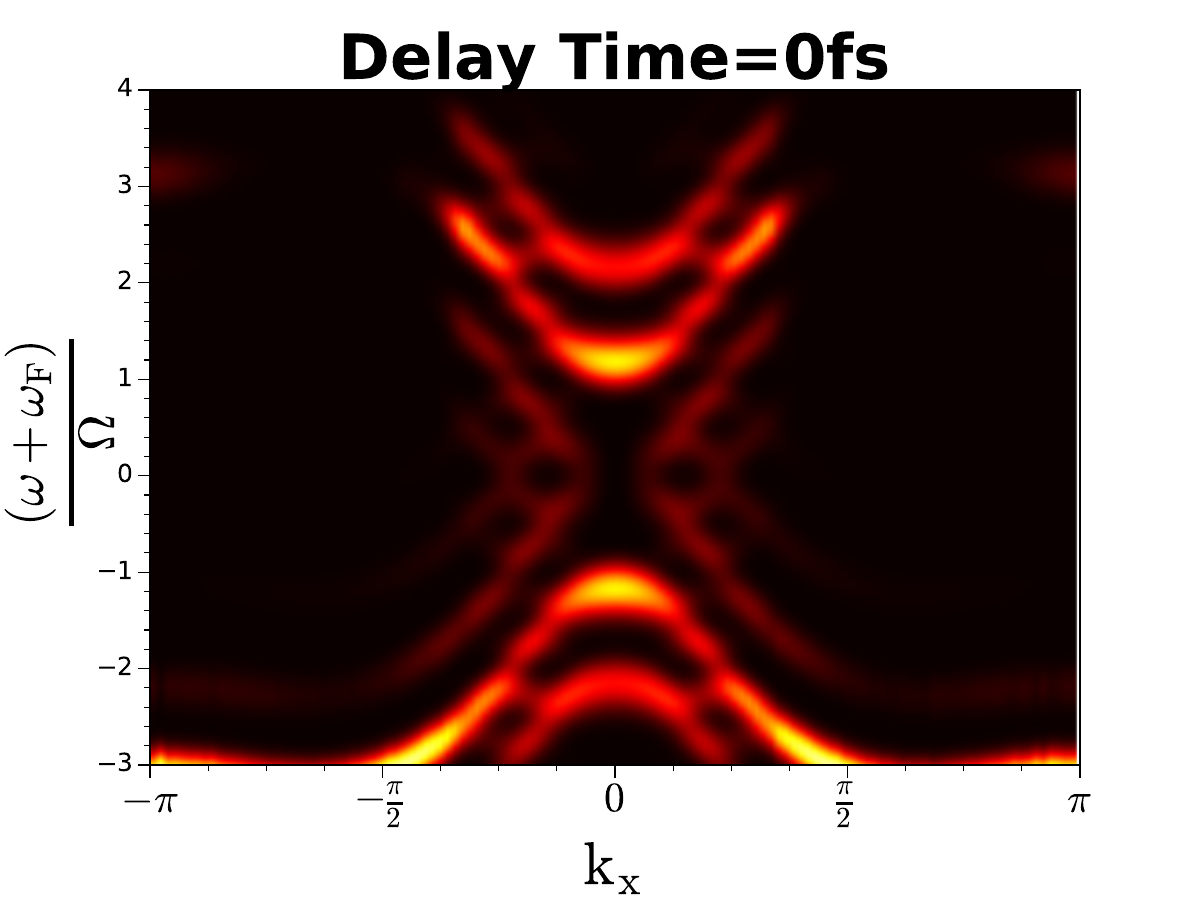}
 	\includegraphics[width = 0.8\columnwidth]{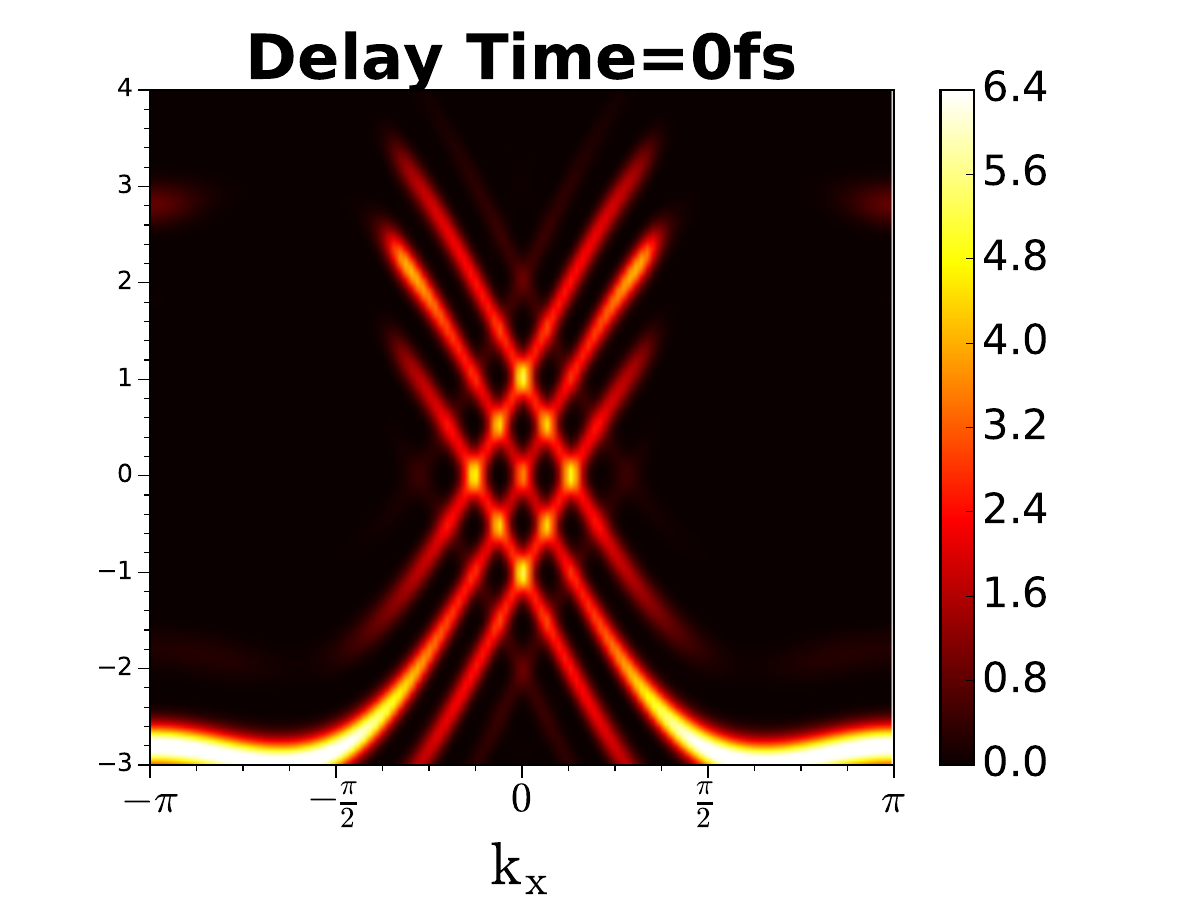}
 	\caption{{\small A snapshot of the time resolved ARPES photocurrent at delay time zero calculated within the lattice model of Eq.~(\ref{eq:Marchand}), with circularly polarized light (left) and linearly polarized light (right). The lattice momentum $k_x$ is varied over the the full range while $k_y$ is kept fixed at zero.  The parameters are chosen such that a Dirac cone appears at equilibrium:  $\lambda = 2t$ and $\mu = 4t$.  The light pulse parameters are the same as before.   
 	}}\label{fig:Marchand}
 \end{figure*}
 \end{widetext}

In addition to the above continuum model we have also used the lattice model of a TI developed by Marchand and Franz\cite{Marchand} to look at the behaviour at the Dirac point. The equilibrium Hamiltonian is given by:
\begin{eqnarray}\label{eq:Marchand}
h_{\kv} &=& \begin{pmatrix} \xi_{\kv} & M_{\kv}\\ M_{\kv} & -\xi_{\kv} \end{pmatrix} \nonumber \\
\xi_{\kv}&=&2\lambda \left( \sin(k_x)\sigma_y -\sin(k_y)\sigma_x \right) \nonumber\\
M_{\kv} &=& -2t\left(\cos(k_x)+\cos(k_y)\right) -\mu 
\end{eqnarray}
We calculate the photocurrent again, in the framework of this model.  While the band curvature away from the Dirac point it definitely apparent, the behaviour at the Dirac point is essentially the same as in the linearized model.  Fig.~\ref{fig:Marchand} is a snapshot of the time result ARPES spectrum at the center of the pump pulse, for $k_y=0$ and both circularly and linearly polarized light.
 \end{document}